\documentclass[a4paper,10pt]{article}
\newcommand{\pdiff}[2]{\ensuremath{\frac{\partial #1}{\partial #2}}} 
\newcommand{\phia}{\ensuremath{\phi_\alpha}}
\usepackage[english]{babel}
\usepackage[utf8x]{inputenc}
\usepackage{graphicx}
\usepackage{caption}
\usepackage[labelformat=simple]{subcaption}

\usepackage{mathtools}
\usepackage{amsfonts}
\usepackage{xfrac}
\usepackage{siunitx}
\usepackage{placeins}
\usepackage{tikz}
\usepackage{pgfplots}
\usetikzlibrary{calc}
\usetikzlibrary{backgrounds}
\usepgflibrary{shapes}
\usetikzlibrary{through}
\usetikzlibrary{intersections}
\usetikzlibrary{matrix}
\usepackage{multirow}
\usepackage{nameref}
\usepackage{url} % srep doesn't like hyperref
\usepackage{cleveref}
\crefname{figure}{Fig.}{Figs.}
\Crefname{figure}{Figure}{Figures}

\crefname{table}{Table}{Tables}
\Crefname{table}{Table}{Tables}

\usepackage{xifthen}

\usepackage{esdiff}
\usepackage{ifthen}

\usepackage{booktabs}
\usepackage{nicefrac}

\usepackage[
]{todonotes}
\presetkeys{todonotes}{inline}{}
\usepackage{footmisc}

\usepackage[singlespacing]{setspace}

\newcommand{\Vap}{\ensuremath{\mathrm{V}}}

\newcommand{\basec}[1]{\textcolor[HTML]{008080}{#1}}
\newcommand{\Dsrc}[1]{\textcolor[HTML]{800080}{#1}}
\newcommand{\Dsrrc}[1]{\textcolor[HTML]{FFA500}{#1}}
\newcommand{\Dgbrc}[1]{\textcolor[HTML]{008000}{#1}}
\newcommand{\DgbrDsrc}[1]{\textcolor[HTML]{0000FF}{#1}}

\newcommand{\cor}[1]{\textcolor{black}{#1}}

\newcommand{\grad}[1]{\ensuremath{\nabla{#1}}}

\newcommand{\ha}{\ensuremath{h_\alpha}}

\newcommand{\contact}{\ensuremath{\mathrm{C}}}

\newdimen\CdotAxis
\newcommand*{\CdotAux}[3]{%
  {%
    \settoheight\CdotAxis{$#2\vcenter{}$}%
    \sbox0{%
      \raisebox\CdotAxis{%
        \scalebox{#1}{%
          \raisebox{-\CdotAxis}{%
            $\mathsurround=0pt #2#3$%
          }%
        }%
      }%
    }%
    \dp0=0pt %
    \sbox2{$#2\bullet$}%
    \ifdim\ht2<\ht0 %
      \ht0=\ht2 %
    \fi
    \sbox2{$\mathsurround=0pt #2#3$}%
    \hbox to \wd2{\hss\usebox{0}\hss}%
  }%
}

\def\addlegendimage{\csname pgfplots@addlegendimage\endcsname}

\DeclareSIUnit{\molpc}{mol\text{-}\%}

\title{Revealing process and material parameter effects on densification via phase-field studies}
\usepackage{authblk}
\author[1,2*]{{Marco Seiz}}

\author[2]{{Henrik Hierl}}
\author[1,2,3]{{Britta Nestler}}
\author[4]{{Wolfgang Rheinheimer}}
 
\affil[1]{{Institute for Applied Materials, Karlsruhe Institute of Technology, Stra\ss{}e am Forum 7, 76131 Karlsruhe, Germany}}
\affil[2]{{Institute of Nanotechnology, Karlsruhe Institute of Technology, Hermann-von-Helmholtz-Platz 1, 76344 Eggenstein-Leopoldshafen, Germany}}
\affil[3]{{Institute for Digital Materials,  Karlsruhe University of Applied Sciences, Moltkestr. 30, 76133 Karlsruhe, Germany}}
\affil[4]{{Institute for Manufacturing Technology of Ceramic Components and Composites, University of Stuttgart, Allmandring 7B, 70569 Stuttgart, Germany }}
\affil[*]{{corresponding author: marco.seiz@partner.kit.edu}}

\usepackage[square,numbers,sort&compress]{natbib}

\providecommand{\keywords}[1]
{
  \small	
  \textbf{\textit{Keywords:}} #1
}
\begin{document}

\maketitle

\begin{abstract}
% \todoMarco{go over plots again + spacing, check supmat}
% \todoMarco{ctrl-f supatmurl before resub because url doesn't expand}
% \todoMarco{reupload supmat after all is done, old README.txt}
% \todoMarco{change corresponding to partner mail}
Sintering is an important processing step in both ceramics and metals processing.
The microstructure resulting from this process determines many materials properties of interest.
Hence the accurate prediction of the microstructure, depending on processing and materials parameters, is of great importance.
The phase-field method offers a way of predicting this microstructural evolution on a mesoscopic scale.
The present paper employs this method to investigate concurrent densification and grain growth and the influence of stress on densification.
Furthermore, the method is applied to simulate the entire freeze-casting process chain for the first time ever by simulating the freezing and sintering processes separately and passing the frozen microstructure to the present sintering model.

% \todoMarco{rewrite}
% The resulting microstructure after the sintering process determines many materials properties of interest.
% In order to understand the microstructural evolution, simulations are often employed.
% One such simulation method is the phase-field method, which has garnered much interest in recent decades.
% However, the method lacks a complete model for sintering, as previous works could show unphysical effects and the inability to reach representative volume elements.
% Thus the present paper aims to close this gap by employing molecular dynamics and determining rules of motion which can be translated to the phase-field model.
% The resulting phase-field model is shown to be representative starting from particle counts between 97 and 262 and contains the qualitative correct dependence of sintering rate on particle size.
\end{abstract}

\section*{Introduction}
% \todoMarco{rewrite}

Many products being used today, from the humble coffee cup to solar cells, undergo the sintering process at some point in their processing chain.
While the precise materials properties of a coffee cup might not be particularly important barring liquid leakage, the electrical contacting of solar cells is important for their efficiency\cite{Schubert2006,Bucherl2013,Schube2018}.
Furthermore, since different materials are being co-sintered, solar cells also face the problem of warping\cite{Brown2009}, similar to solide oxide fuel cells.
\cor{Finally, sintering is also usually used in the freeze-casting process chain to improve the mechanical properties of the structure.
The porosity produced during the freeze-casting process is of two types:
Macropores, aligned to the freezing direction, and micropores located in the walls of particles formed during the process.
The goal of sintering this structure is to close the micropores, giving mechanical strength to the body, while keeping the macropores and thus achieving a highly porous, low-density but still sufficiently strong material.
For example, this kind of material is of interest for tissue engineering scaffolds, in which the lack of small, interconnected porosity avoids the slowing down of cell proliferation into these regions\cite{Deville2017}.
Furthermore, since the porosity can be made gradiated and is highly anisotropic, this also favors penetration of tissue into the freeze-cast scaffold\cite{Li2013}.}\\
This requirement for controlling the microstructure, resulting materials parameters and dimensions is shared across many sintered products.
Hence predicting the microstructure and its properties as well as the sintered dimensions is of great importance.
Key properties of the microstructure are its density, grain and pore size as well as shape distributions, based on which further properties such as resistivity can be predicted.

In the past two decades, the phase-field method has been used by many researchers\cite{kazaryan1999generalized,Wang2006,kumar2010phase,biswas2016study,Hoetzer2019,Abdeljawad2019,Dzepina2019,Termuhlen2021,Shi2021} to investigate the sintering process.
More recent works have pointed out several weaknesses of these models\cite{Seiz2022,Seiz2023a,Yang2022b}, which were rectified subsequently\cite{Yang2022b,Seiz2023a,Seiz2023b}.
Based on these works, the present paper aims to elucidate several factors of interest in the solid-state sintering process:
External pressure is often applied to speed up densification:
Capillary pressure acts as a driving force for sintering and applying external pressure enhances this.
Furthermore, the entire sintering process is fundamentally one of mass transport, with multiple concurrently active transport pathways such as grain boundaries (GBs) and surfaces.
The ratio of their contribution plays a significant role, since densification is due to vacancy absorption on grain boundaries, with surface diffusion possibly acting counter to this mechanism.
As elevated temperatures are usually employed for sintering, grain growth is another matter of concern.
This not only increases the mean grain size, reducing e.g. mechanical strength, but also allows for porosity to detach from boundaries and be trapped within grains.
Once this occurs, these pores cannot be eliminated without pressure on processing timescales.
Hence the combination of grain mobility, as well as species mobility along interfaces, is another factor of interest.
Finally, the mass distribution and pore anisotropy of a green body can cause anisotropic densification\cite{Lichtner2018,Farhangdoust2013}.

The effect of pressure will be investigated first, as it is a comparatively simple investigation.
Since sintering and creep are closely related processes, a short excursion to creep is done as well to show the versatility of the model.
Next, the materials properties of diffusivity and grain mobility will be varied, effectively representing sintering at different holding temperatures or with appropriate dopings.
Special attention is paid to how different transport and growth mechanisms affect each other's influence on the microstructural evolution.
In order to investigate the effect of green body geometry, a previously generated particle structure of the freeze-cast process is sintered, showing the experimentally observed anisotropic shrinkage of freeze-cast structures.
The results of these investigations are presented first, with the employed phase-field model being succinctly described in \nameref{sec:model} and in further references therein.

The materials parameters are those of \cref{tab:params} unless mentioned otherwise.
These parameters approximate copper at around $\SI{700}{K}$.
\cor{Copper is chosen as an example material as much experimental data of its sintering is available.
Furthermore, since it is a metal, many of the complexities introduced when sintering ceramics, such as charged defects and space charge zones across GBs, can be safely ignored.
Applications using sintered copper include microelectronic applications as well as foams used for heat exchangers, which can be produced via freeze-casting followed by sintering\cite{Ramos2012}.}
\cor{Simple scalings of the diffusion and mobility values are taken to approximate different holding temperatures, such as to isolate the effects of the individual parameters.}

\section*{Results \& Discussion}

\begin{figure}[h]
 \centering
 \includegraphics[width=0.8\textwidth]{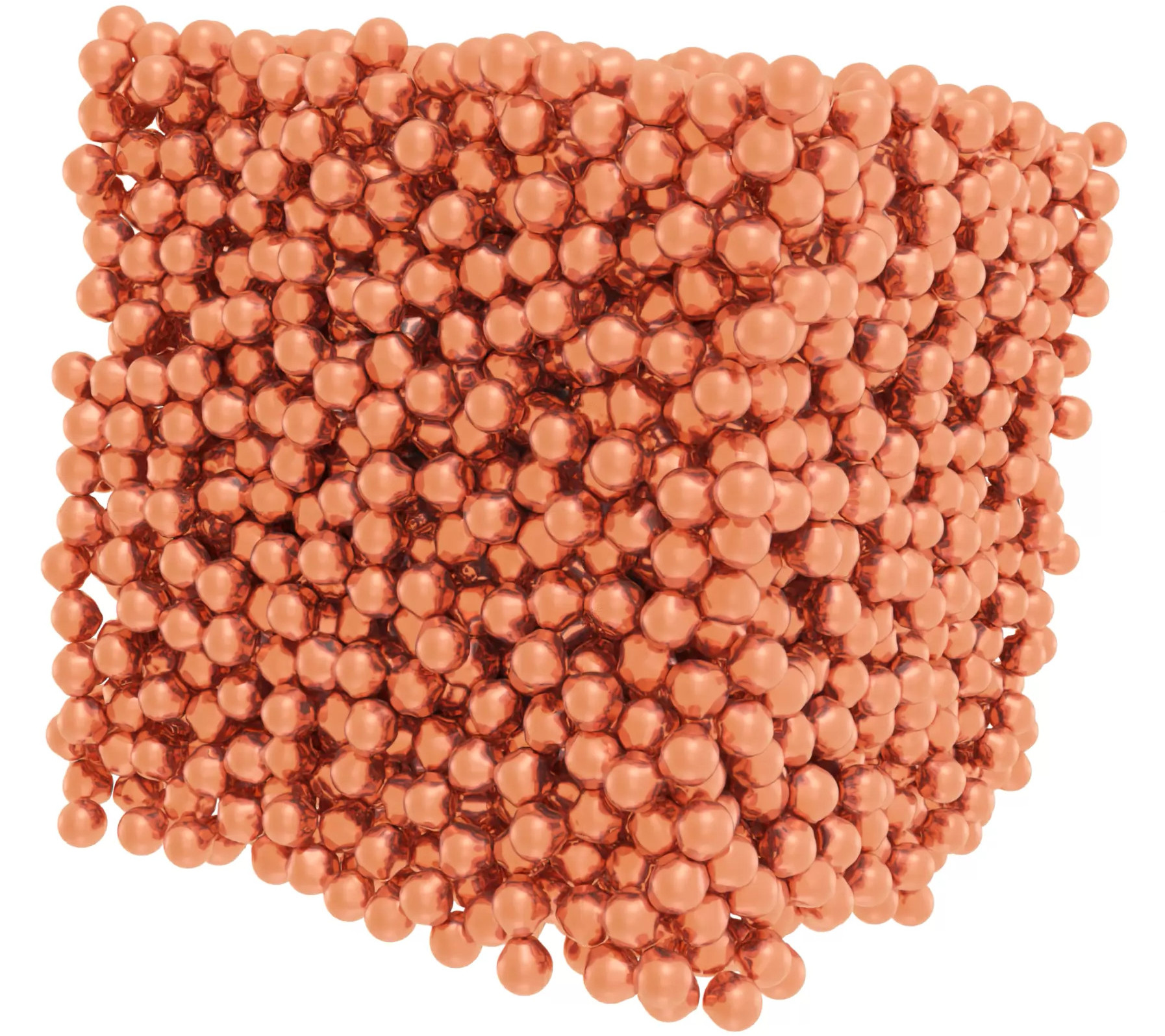}
 \caption{\cor{The initial three-dimensional green body used for simulations, except for simulations concerning creep and freeze-casting, is shown.
 The overall shape is a cube, with spheres packed into the cube based on a larger packing from a discrete element method simulation; for details, see \cite{Seiz2023b}.
 The grains are shown as a copper-like material.}}
 \label{fig:initial}
\end{figure}

Based on prior work \cite{Seiz2023b}, the density evolution of 3D phase-field simulations of solid state sintering \cor{does not change substantially after more than 262 particles are contained within the green body, given that it starts with a uniform grain size and holds this size over the simulation run.}
If not mentioned otherwise, we shall employ a packing with 3445 spheres of radius $r=\SI{12}{nm}$ in a $\SI{400}{nm^3}$ box for the initial conditions and observe how its evolution changes when various processing and materials parameters are changed.
\cor{A small grain size is employed for computational efficiency, but as long as Herring's scaling law holds as shown in the supplementary material, the present results can be mapped onto coarser grain sizes.}
\cor{The initial green body of this packing is depicted in \cref{fig:initial} representing the grain structure}.
The obtained densification should be \cor{independent of domain size} based on \cite{Seiz2023b}, but the same need not hold for the grain size evolution since the number of remaining grains can become quite small.
Additionally, the influence of green body geometry is analyzed by employing a freeze-cast structure obtained with the model of \cite{Seiz2021b} and sintering it.

\subsection*{Sintering under stress}

The first foray will consider the influence of stress on sintering, as it is quite simple but also important for densification.
Pressure is often applied during sintering in order to either speed up densification or to remove isolated/detached porosity by forcefully dissolving the contained gas into the material.
In this section the former effect is investigated by considering the effect of the applied stress on the equilibrium concentration of vacancies on grain boundaries.
This effect can be easily included in the model of \cite{Seiz2023b} as shown in the modelling section. %\nameref{sec:model}.
%\cref{sec:model}.

\begin{figure}[h]
\centering
     \begin{subfigure}[]{0.9\columnwidth}
        \includegraphics[width=\textwidth]{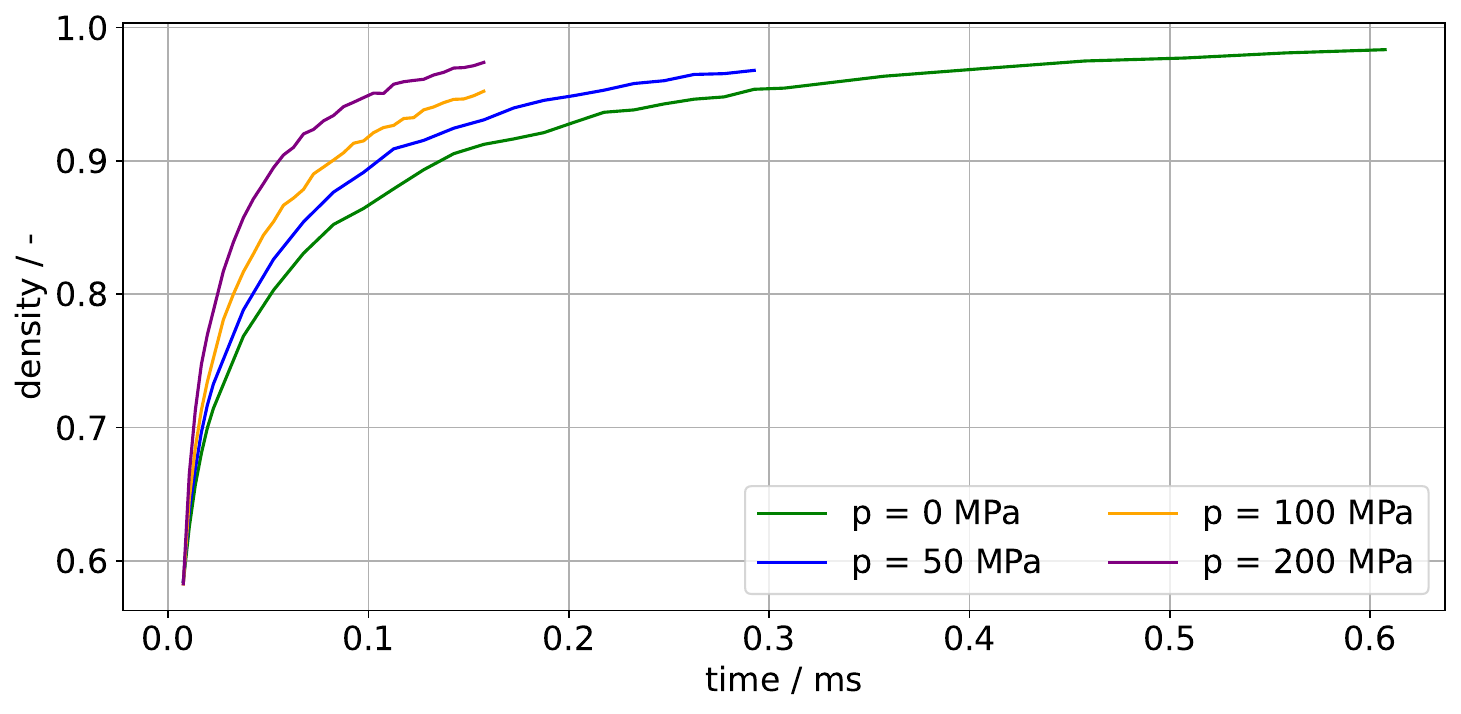}
        \caption{time evolution of density for various pressures}
 \label{fig:densi-md-press}
    \end{subfigure}
~
    \begin{subfigure}[]{0.9\columnwidth}
    \centering
        \includegraphics[width=\textwidth]{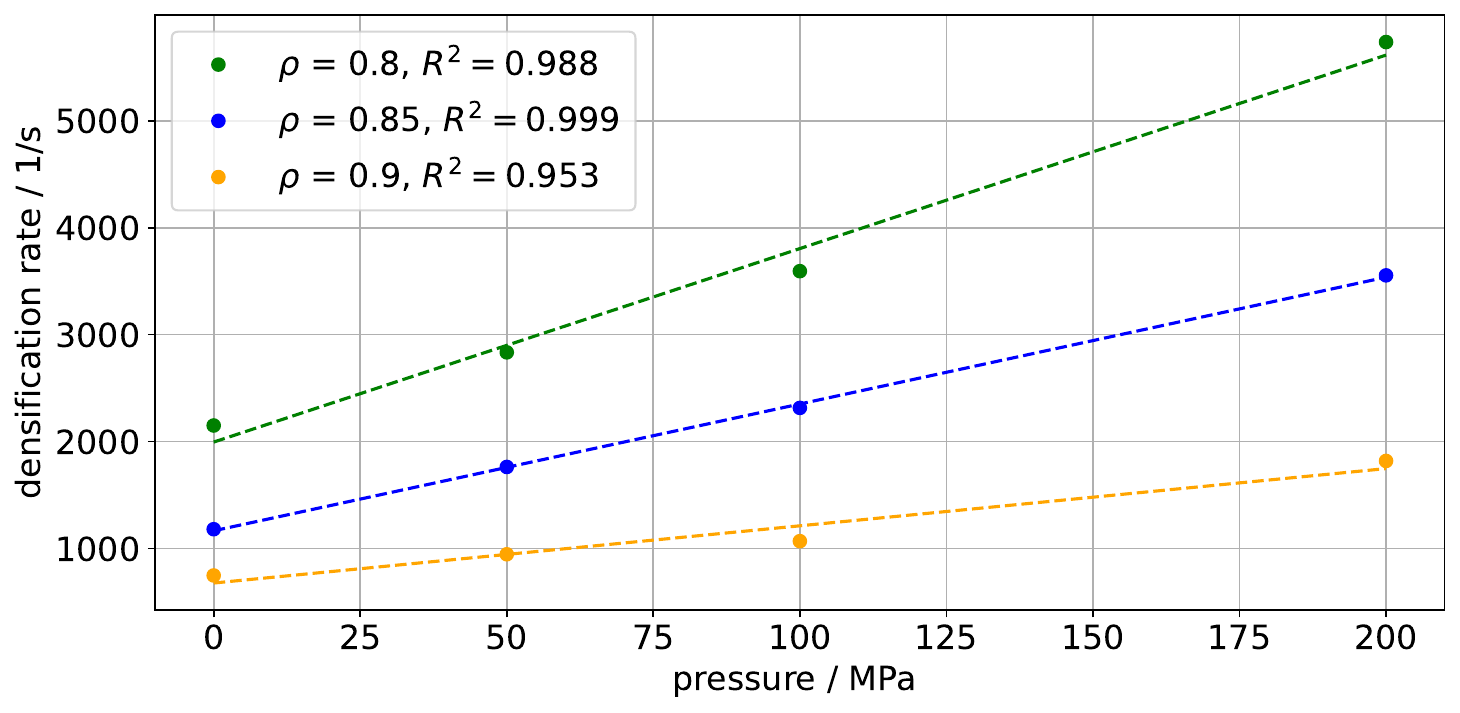}
        \caption{pressure influence on densification rate}
        \label{fig:densirate-pressure}
    \end{subfigure}
 \caption{The influence of pressure on the density evolution is depicted in (a), with its influence on densification rate depicted in (b).
 With rising pressure, quicker densification is achieved.
 The dashed lines in (b) indicate linear fits with coefficient of determination $R^2$ given in the legend.
 There is a roughly linear relationship between pressure and densification rate as also experimentally observed by \cite{Harmer1980}.
 }
 \label{fig:press-influence}
\end{figure}

The pressures $p \in \{0, 50, 100, 200\} \si{MPa}$ will be considered. 
The scale of the pressure is based on the magnitude of the capillary pressure $O(p_c) \sim  O(\frac{\gamma_{s}}{r}) \approx \SI{100}{MPa}$ with the surface energy $\gamma_s$ and the particle radius $r$.
The resulting density over time is shown in \cref{fig:densi-md-press}, with increasing densification rate with higher pressures.
In order to determine the qualitative relationship between pressure and densification, the densification rate of the individual simulations is calculated at test densities of $\rho \in \{ 0.8, 0.85, 0.9 \}$.
For this, the numerical densification rate, i.e. the slope in \cref{fig:densi-md-press} is calculated, followed by an interpolation across its density space, which then allows the calculation of densification rates at any density within the valid density range.
This yields \cref{fig:densirate-pressure} which shows a roughly linear relationship between densification rate and pressure, as also observed by experiments \cite{Harmer1980}.
Classical theory e.g. due to Coble \cite{Coble1970} also predicts a linear relationship between external pressure and densification rate.
Hence this simple addition of pressure compares well against both experiment and theory.

\begin{figure}[h]
\centering
 \includegraphics[width=0.9\columnwidth]{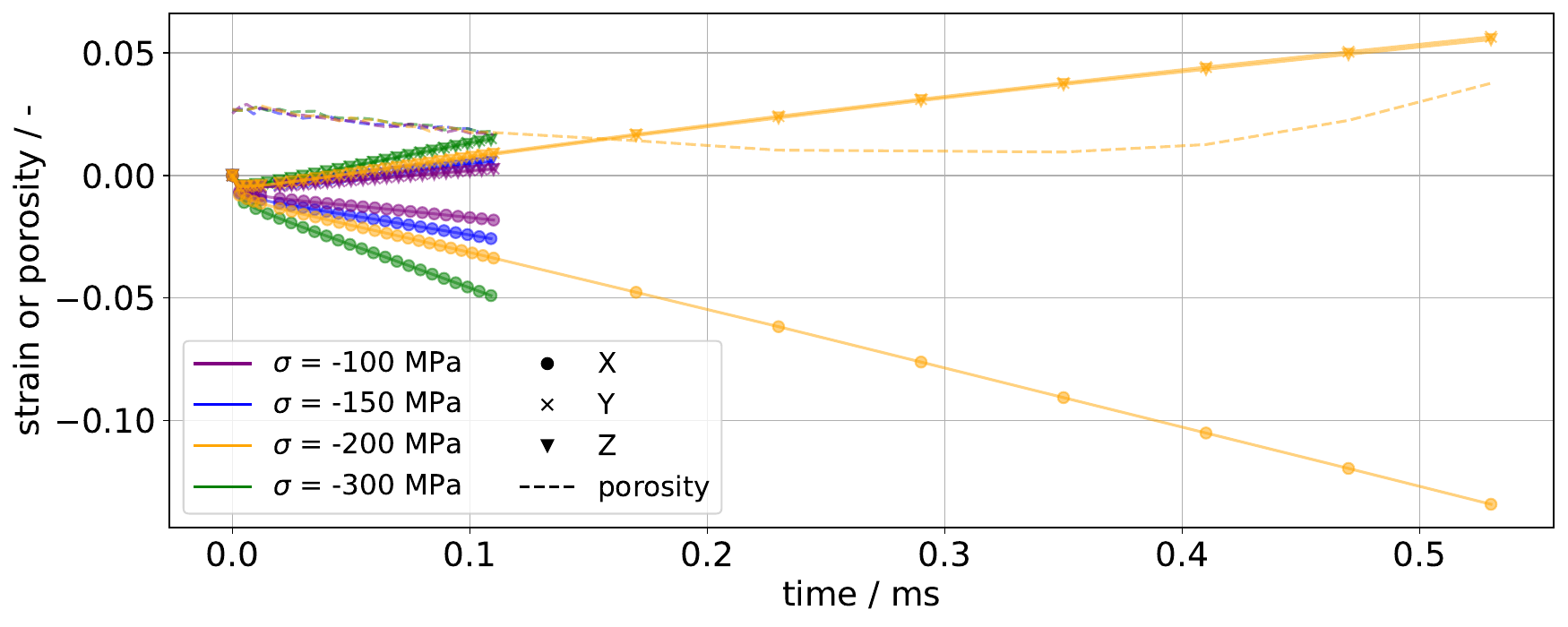}
 \caption{Strain and porosity evolution for several creep simulations, with tensile stress being applied in the $X$ direction.
 After an initial transient, a constant strain rate is observed, characteristic of secondary creep.
 }
 \label{fig:creep-strain}
\end{figure}

As sintering and creep are intimately related processes, any model of solid-state sintering should also be able to approximate creep.
The present model can achieve this by setting the stress to be uniaxial, which will cause vacancy generation on grain boundaries aligned with the tensile stress direction.
Four tensile tensile stresses $\sigma \in \{ -100, -150, -200, -300\} \si{MPa}$ in the $X$ direction are considered, with $\sigma = \SI{-200}{MPa}$ being run longer than the others to show porosity increase.
The simulations are conducted by starting from an almost dense (97\%) sintered body produced by prior simulations.
The evolution of normal strain in the coordinate directions as well as porosity are shown in \cref{fig:creep-strain}:
The samples lengthen (negative strain) in the $X$ direction and shrink in the $Y$ and $Z$ directions, with the $Y$ and $Z$ strains effectively overlapping.
After an initial transient, the strain is observed to be a linear function of time as expected of secondary creep.
The steady strain rate is observed to be linearly dependent on the applied stress, as is expected of Coble creep \cite{Coble1963}.
Porosity is reduced at first, but with sufficient time, the porosity rises and would eventually lead to failure.
This suggests that the model extension successfully models the stress dependence of Coble creep.

%maybe put a rendering of porosity nucleated on the right/left

\FloatBarrier

\subsection*{Coupling grain growth and densification}

\begin{figure}[]
\includegraphics[width=\textwidth]{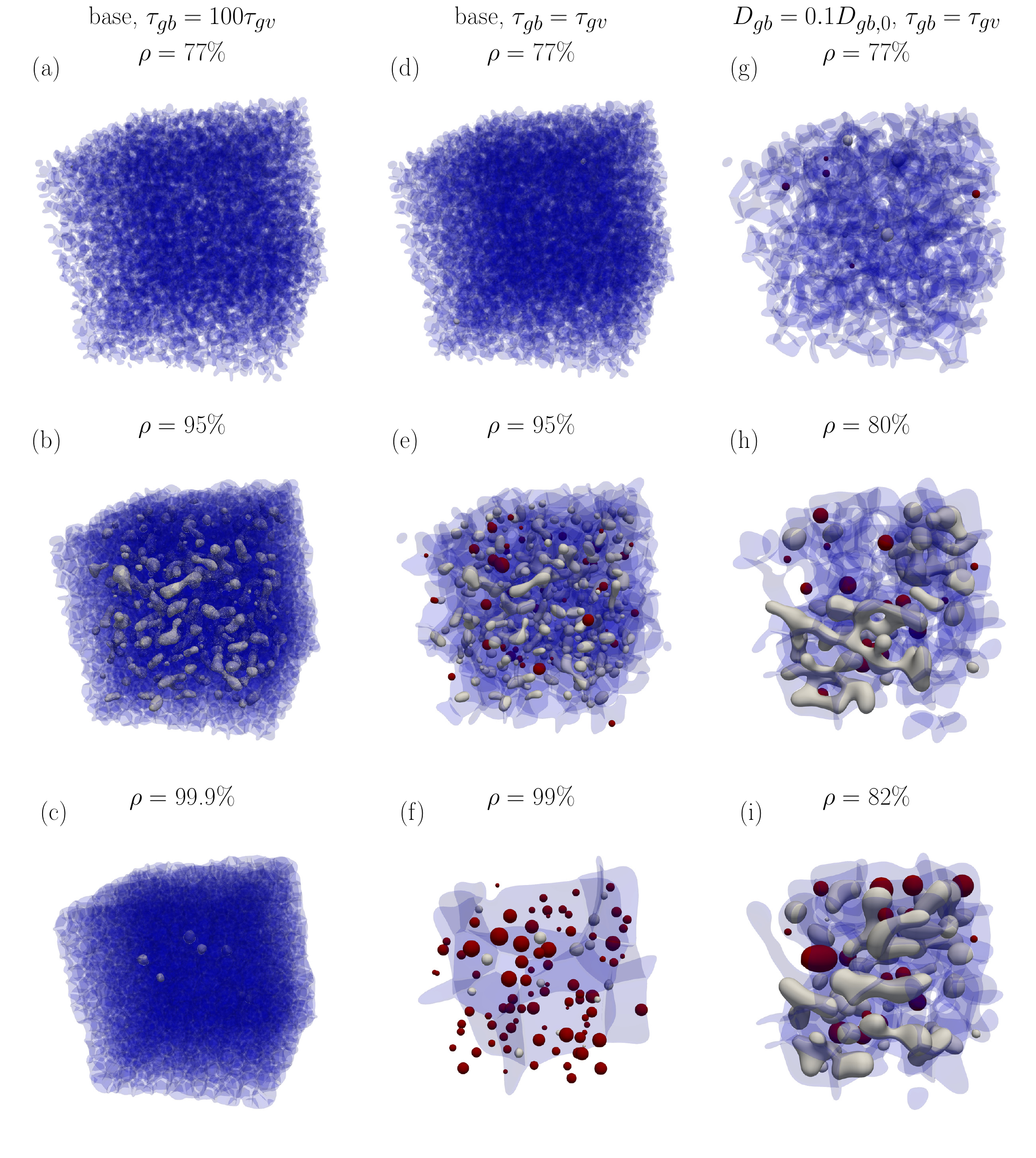}
 \caption{Comparison of simulations with \basec{suppressed grain growth} (left\cor{, a-c}), with \basec{grain growth} (middle\cor{, d-f}) and a simulation with \Dgbrc{grain growth and reduced GB diffusivity} (right\cor{, g-i}).
 The GB network is shown in transparent \textcolor{blue}{blue}, isolated porosity in grey and detached porosity in \textcolor{red}{red}.
 The depicted states were chosen based on their density and can represent different times.}
 \label{fig:visu}
\end{figure}

Densification and grain growth are intimately related and thus the concurrent simulation of both is of paramount importance for quantitative simulation of sintering.
In this section, these coupled processes are investigated by parameter variations of the relevant mobilities.
The mobility of an interface is controlled with the inverse mobility $\tau$, for which high values imply low mobility.
For the grain-vapor interface a small value of $\tau_{gv}$ is chosen such that the evolution is diffusion-controlled.
For the grain-grain interfaces, two different values of the inverse grain mobility $\tau_{gb}$ are chosen, with their values also serving as simulation labels:
Kinetically suppressed grain growth is achieved with $\tau_{gb} = 100\tau_{gv}$ and unsuppressed grain growth with $\tau_{gb} = \tau_{gv}$.

For each of these mobility variations, different diffusion coefficients will be tested.
The base case is that of \cref{tab:params}, with variations on the surface diffusivity $D_s$ and the GB diffusivity $D_{gb}$ by simple factors, with the factors being chosen arbitrarily.
The full set of investigated diffusivity combinations is 
\begin{align}
\{&(\basec{D_{gb,0}, D_{s,0}}), (\Dsrc{D_{gb,0}, 0.1D_{s,0}}), (\Dsrrc{D_{gb,0}, 0.01D_{s,0}}), \nonumber \\
&(\Dgbrc{0.1D_{gb,0}, D_{s,0}}), (\DgbrDsrc{0.1D_{gb,0}, 0.01D_{s,0}})\} \nonumber                                                                                                                                                                                                   
 \end{align}
 with the color scheme being used consistently to identify the simulations.
The base case $\basec{(D_{gb,0}, D_{s,0})}$ is identified explicitly, with departures from it being used as labels with the appropriate color.
The surface to GB diffusion ratio \basec{in the base case of $\frac{D_{s,0}}{D_{gb,0}}$} is about 3, with this being varied from about 1/30 to 30 with the employed factors.

A qualitative, visual comparison of some of these results is given in \cref{fig:visu} showing the GB network (blue) as well as isolated (grey) and detached porosity (red) for selected densities.
It can easily be seen that \basec{in the base case without grain growth} (left), no pores are detached.
For otherwise the \basec{same parameters} (middle), allowing grain growth leads to a significant fraction ($\approx 80\%$) of detached porosity.
Finally, if the \Dgbrc{GB diffusivity is reduced} (right), then densification slows down significantly and within the allotted simulation time the maximum reached density is about 82\%, whereas both simulations with quick GB diffusion achieved at least $99\%$ density.
As an exemplary quantification, the time to reach $80\%$ density is increased by a factor of 33 even though the GB diffusion was only reduced by a factor of 10.
From these images it should also be clear that even at the same density, microstructures need not be comparable.
Rather, these are the complex product of the temporal interplay of densification, pore destabilization and grain growth, with a single variable being unable to capture the full picture.
Videos of the time evolution, also showing a kind of Plateau-Rayleigh instability of isolated pore channels, are deposited with the supplementary material \cor{(\url{https://doi.org/10.5281/zenodo.8263532})}.

\begin{figure}[h!]
\begin{center}
     \begin{subfigure}[]{0.9\columnwidth}
    \centering
        \includegraphics[width=\textwidth]{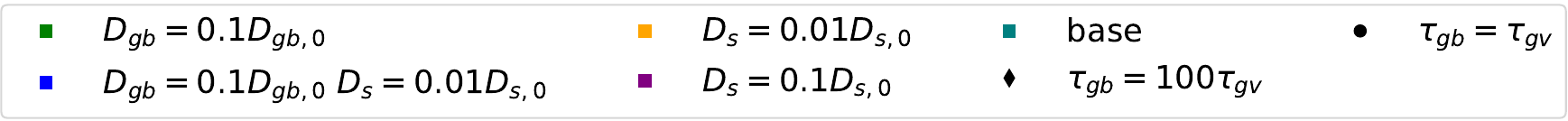}
    \end{subfigure}
     \begin{subfigure}[]{0.9\columnwidth}
    \centering
        \includegraphics[width=\textwidth]{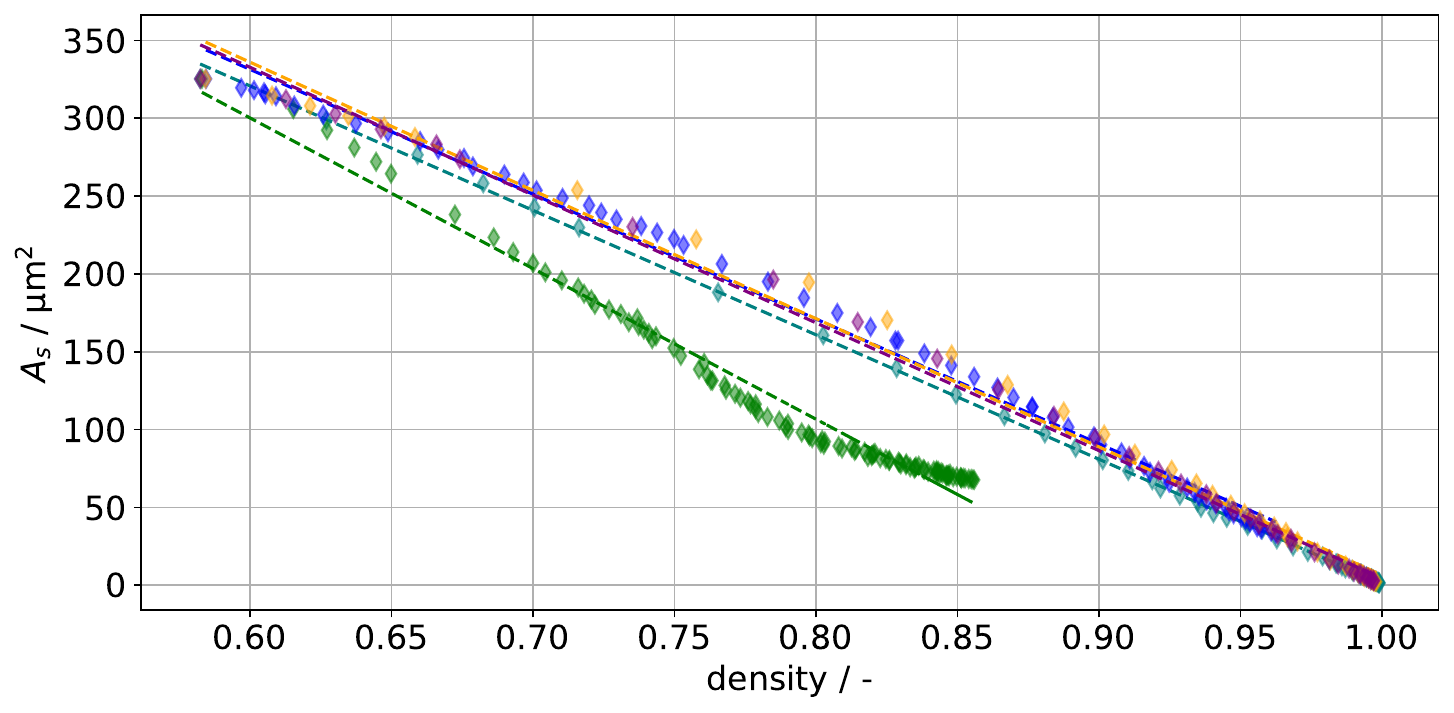}
        \caption{suppressed grain growth}
        \label{fig:surf-densi-nogg}
    \end{subfigure}
        \begin{subfigure}[]{0.9\columnwidth}
    \centering
        \includegraphics[width=\textwidth]{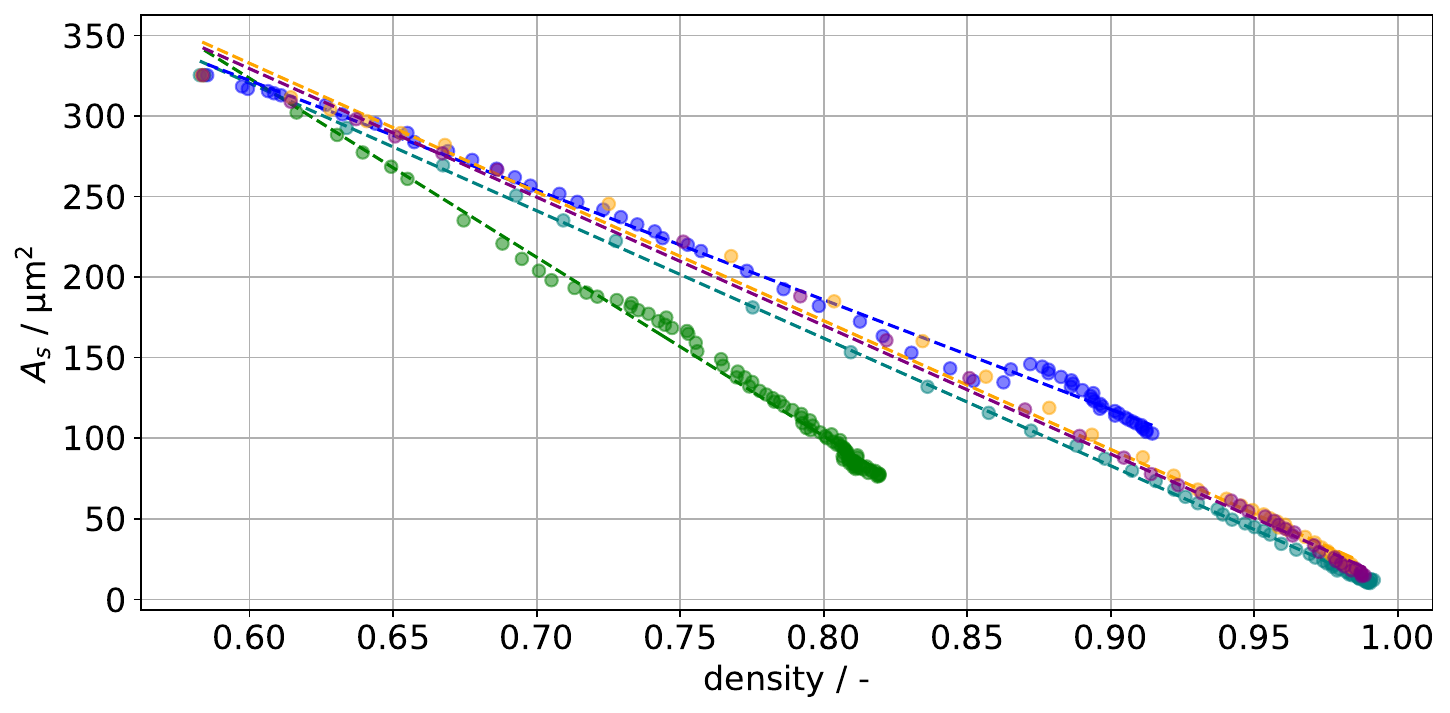}
        \caption{unsuppressed grain growth}
        \label{fig:surf-densi-gg}
    \end{subfigure}
 \caption{The relationship between surface area and density is observed to be roughly linear, with the dashed lines indicating best fit linear functions.
 }
 \label{fig:surf-densi}
 \end{center}
\end{figure}

Let us start the quantitative investigation by considering the evolution of surface area with density, as the reduction of surface area is the driving force for densification.
It is often observed \cite{Wakai2014,Schleef2014,German2016} that surface area and density are linearly related during sintering.
This relationship is demonstrated in \cref{fig:surf-densi} for simulations with and without suppressed grain growth separately.
The legend employed will be the same for all following plots, with colors indicating variations in diffusivity and symbols indicating suppressed grain growth (diamond) or unsuppressed grain growth (circle).
The main point however is that regardless of the employed parameters, a linear relationship is observed.
The linear fits' (dashed lines) coefficient of determination $R^2>0.98$ also shows this in a quantitative manner.
Therefore, the model's densification behaviour is in qualitative accordance with experiments.
Local deviations, such as for the \DgbrDsrc{blue} circles, are mostly due to either a change of sintering stage or the onset of grain growth.
Hence next the density and grain size evolution will be considered.

\begin{figure}[]
\centering
     \begin{subfigure}[]{0.9\columnwidth}
    \centering
        \includegraphics[width=\textwidth]{images/legend}
    \end{subfigure}
     \begin{subfigure}[]{0.9\columnwidth}
        \includegraphics[width=\textwidth]{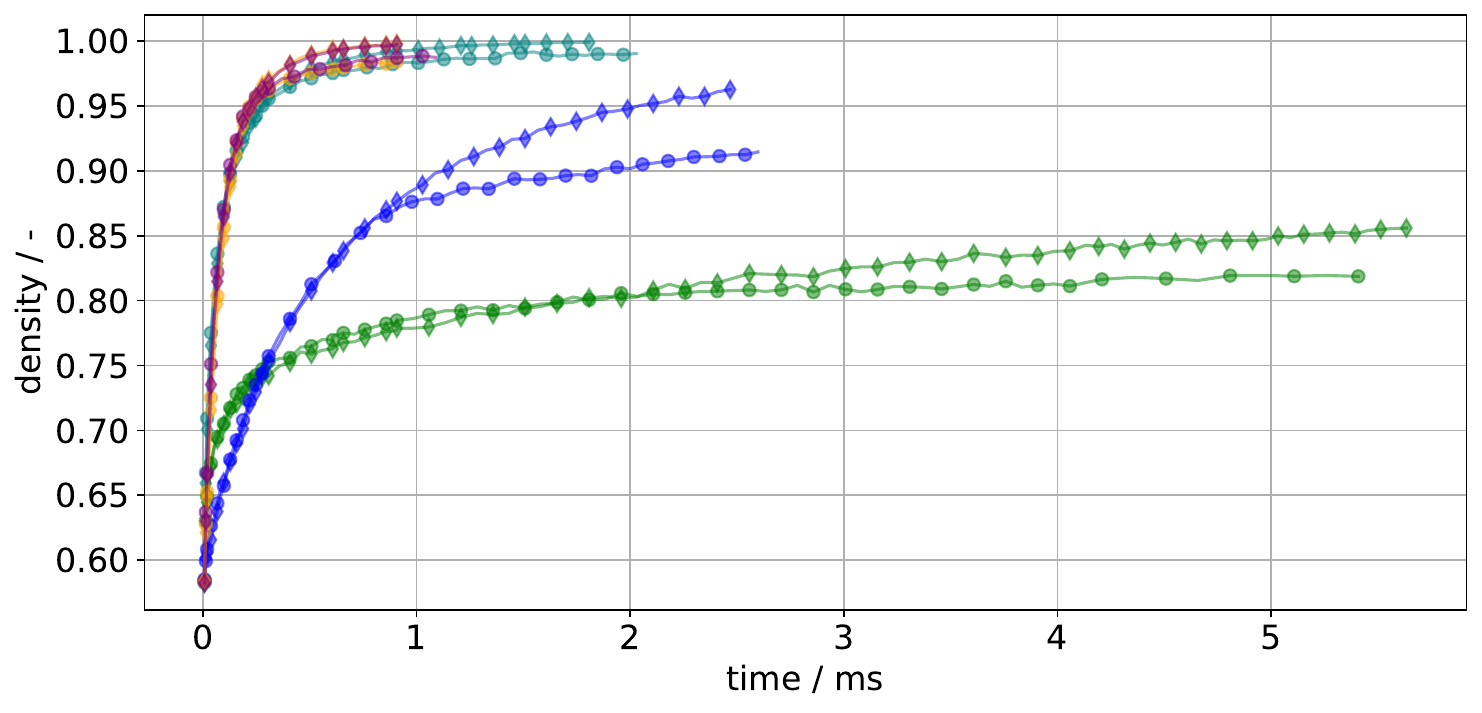}
        \caption{density evolution}
 \label{fig:densi-Dvar}
    \end{subfigure}
~
    \begin{subfigure}[]{0.9\columnwidth}
    \centering
        \includegraphics[width=\textwidth]{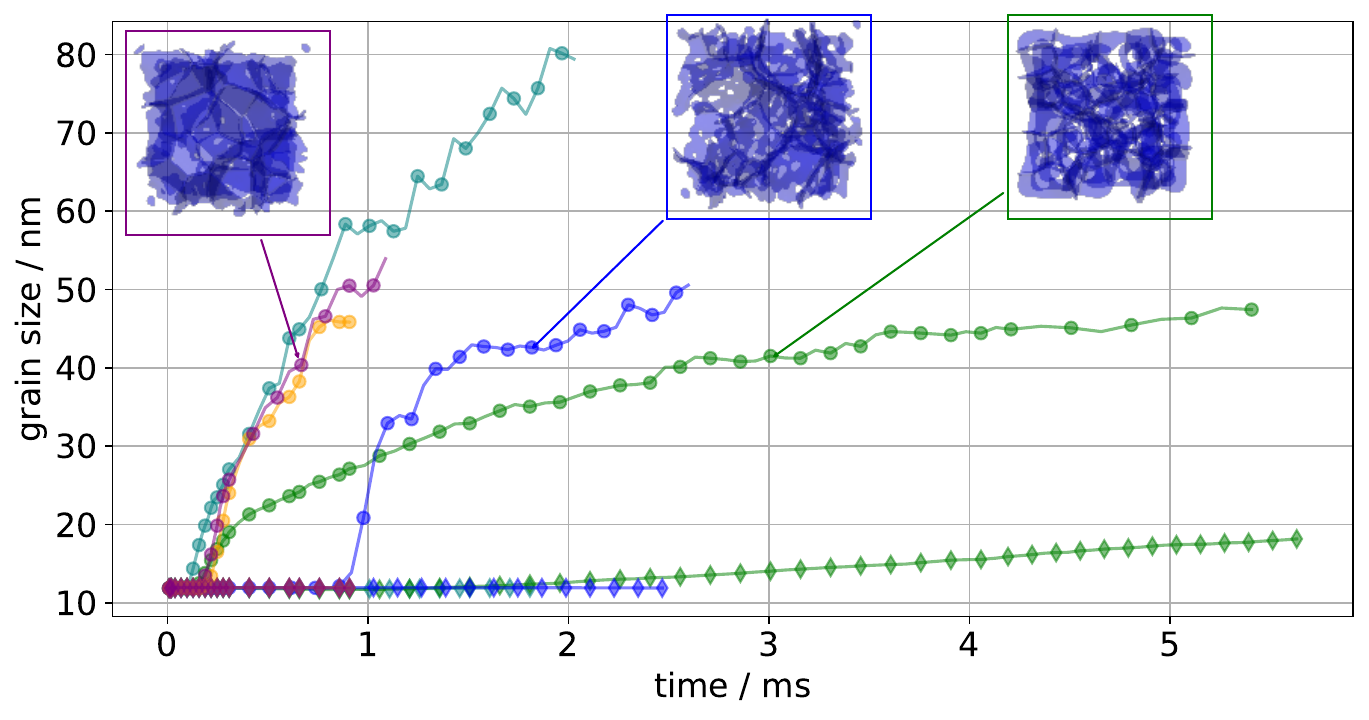}
        \caption{grain size evolution} 
        \label{fig:gg-Dvar}
    \end{subfigure}
    \begin{subfigure}[]{0.9\columnwidth}
    \centering
        \includegraphics[width=\textwidth]{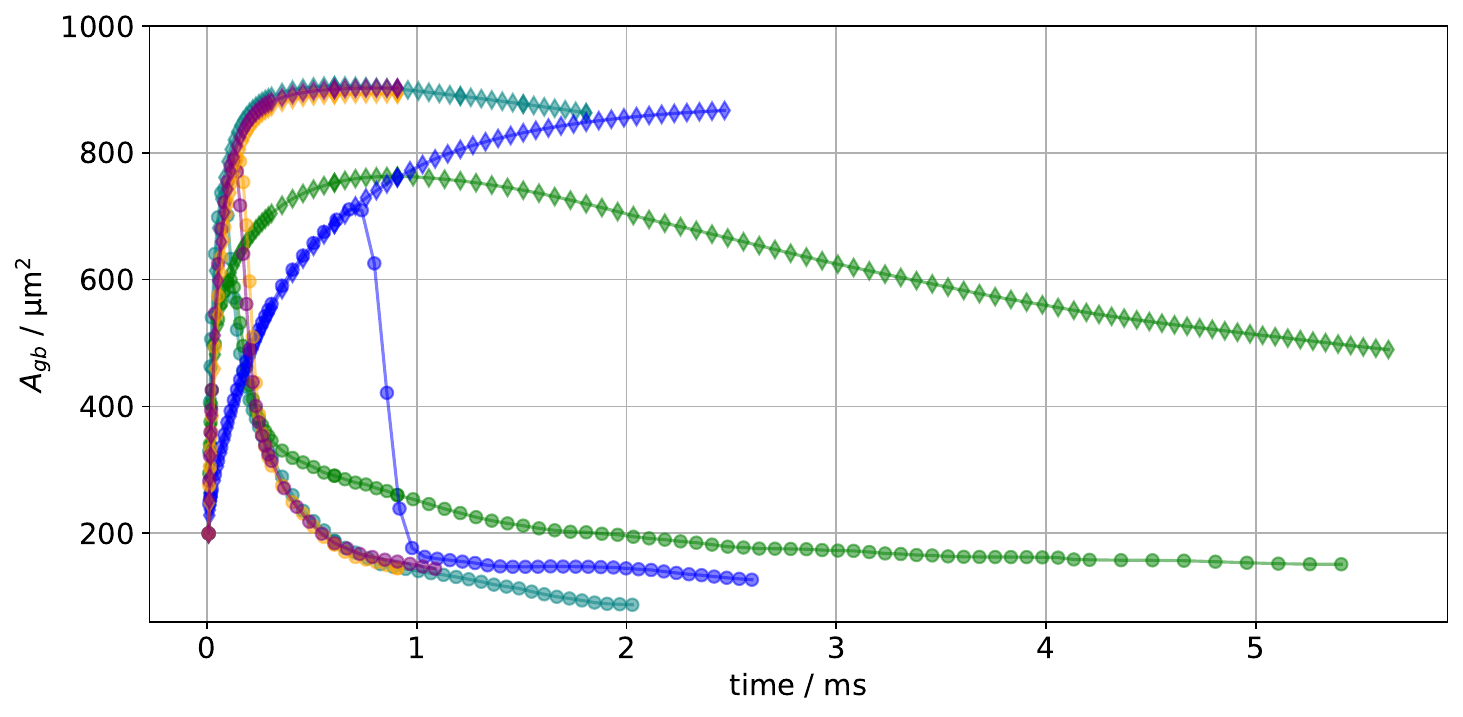}
        \caption{GB area evolution} 
        \label{fig:gbarea}
    \end{subfigure}
 \caption{The influence of grain growth and variable interfacial diffusivity on densification (a), grain size (b) and total GB area $A_{gb}$ (c).
 The GB networks (blue) for three simulations are shown as insets in (b) for similar grain sizes, with white space within the image indicating porosity.
 }
 \label{fig:gg-dvar-results}
\end{figure}

The simulation measurements in terms of density $\rho$, grain size $G$ and GB area $A_{gb}$ are collected in \cref{fig:gg-dvar-results}.
Focus first on the diamonds in \cref{fig:densi-Dvar}:
These represent the density of simulations with suppressed grain growth.
The \basec{base case (teal)}, as well as the simulations in which only surface diffusion $D_s$ is reduced (\Dsrc{purple}, \Dsrrc{orange}), have essentially the same densification behaviour.
This is due to vacancy absorption happening fast enough that surface diffusion contributes relatively little to neck growth compared to vacancy absorption.
There is a small effect of reduced densification rate with reduced surface diffusion up to $\approx\SI{90}{\percent}$ density, after which the densification rate with reduced surface diffusion is larger.
This density-dependent influence of the surface diffusion is likely due to the inversion of the surface mass flux observed by Luo et al. \cite{Luo2015}.
In contrast, \Dgbrc{decreasing the GB diffusion $D_{gb}$ (green)}  significantly reduces the densification rate throughout the process.
In this case a significant amount of neck growth is due to surface diffusion filling the neck without concomitant vacancy absorption, \cor{which also causes the neck curvature to be lower at the same density, shown in the supplementary material.}
Hence \DgbrDsrc{decreasing the surface diffusion (blue)} for this case does significantly speed up densification ($\approx$ factor 3 less time to 80\% density), though not to the original levels.
The two simulations with reduced grain boundary diffusion (\Dgbrc{green}, \DgbrDsrc{blue}) will also henceforth be called slowly densifying, in contrast to the quick densification exhibited by the remaining simulations.

Focus now on the effect of grain growth on densification by comparing the circular symbols (unsuppressed grain growth) and the diamonds (suppressed grain growth) in \cref{fig:densi-Dvar}:
For the \basec{quickly} \Dsrc{densifying} \Dsrrc{simulations} there is hardly any influence on the density evolution.
This is due to densification happening so quickly that grain growth only really starts past about $90\%$ relative density.
Grain growth however does lead to pores detaching from the GBs, which at the end of the simulations causes about $1\%$ porosity to remain in a detached state.
As before, \Dgbrc{decreasing the} \DgbrDsrc{GB diffusion} causes a significant slowdown of densification, with grain growth further limiting the achievable density for a fixed time.
Since grain growth has a significant effect on these, it will be discussed next.

The grain size evolution is depicted in \cref{fig:gg-Dvar}.
Comparing the \basec{base case (teal)} with only reduced surface diffusion (\Dsrc{purple}, \Dsrrc{orange}) shows a slight reduction of grain growth via pore drag, since pore motion is limited by surface diffusion.
The effect is not particularly pronounced due to the quick densification removing pores quickly as well.
Pore drag becomes much more evident when comparing the quickly densifying simulations to the slowly densifying ones:
As densification proceeds more slowly, more pores will be present at the same time and hence grain growth is slowed down.
Unexpectedly, when in addition to reducing the \Dgbrc{GB diffusion} the \DgbrDsrc{surface diffusion is reduced} as well, grain growth starts much later in term of time and density.
This is likely due to the sintering neck being formed only slowly in this case (\DgbrDsrc{$D_{gb} = 0.1D_{gb,0}$ $D_{s} = 0.01D_{s,0}$, blue}), which also suppresses grain growth:
The flux of atoms leading to grain growth is the contact area times the flux density.
The flux density is due to curvature differences between grains and will be roughly comparable between simulations prior to the start of grain growth, as the initial conditions are the same.
The contact area evolution differs significantly however, as is shown in  \cref{fig:gbarea} by identifying it with the GB area $A_{gb}$.
The simulation \DgbrDsrc{$D_{gb} = 0.1D_{gb,0}$ $D_{s} = 0.01D_{s,0}$ (blue)} indeed shows much slower GB area growth.
For this particular simulation, once a few grains have achieved a sufficient neck size, these grow rapidly (up to 6 times the mean grain size) until they are slowed down by porosity again, leading to stagnant grain growth for a short interval.
This heavier porosity loading on grain boundaries is shown via the insets in \cref{fig:gg-Dvar}, which show the GB network for similar grain sizes:
Both simulations with reduced GB diffusivity show increased porosity on grain boundaries, with their grain growth hence being more significantly affected by pore drag.

In order to quantify grain growth with respect to experiments, the grain growth law is evaluated by fitting the grain size $G$ data for $G > \SI{13}{nm}$ to power laws of the form $G=At^n$, with the plots showing the results being deposited with the supplementary material.
The grain size filter is employed to fit only the regime where grain growth is taking place.
Exponents ranging from about $\frac{1}{3}$ to $\frac{4}{5}$ are observed; the experimentally observed range is $\frac{1}{4}$ to $\frac{1}{2}$ \cite{Rahaman2003}.
The simulations with \Dgbrc{$D_{gb} = 0.1D_{gb,0}$} show values close to $\frac{1}{3}$ which is often experimentally observed and can be due to pore drag.
The simulations showing $n>\frac{1}{2}$ are likely due to the start of grain growth significantly deviating from the rest of the curve.
For example, if the data is filtered to $G > \SI{20}{nm}$, then these exponents move appreciably closer to $\frac{1}{2}$, whereas \Dgbrc{$D_{gb} = 0.1D_{gb,0}$} achieves $n=\frac{1}{3}$.
Hence filters for larger grain sizes will cut off more of the initial regime, but also include more data in the regime of low grain numbers.
Ideally a larger system containing more grains and hence capable of reaching steady state would have been simulated.
But since the primary goal was investigating densification, a too small number of grains was chosen for the initial conditions.

\begin{figure}[b!]
\centering
\begin{subfigure}[]{0.9\columnwidth}
    \includegraphics[width=\textwidth]{images/legend}
  \end{subfigure}
  \begin{subfigure}[]{0.9\columnwidth}
    \includegraphics[width=\textwidth]{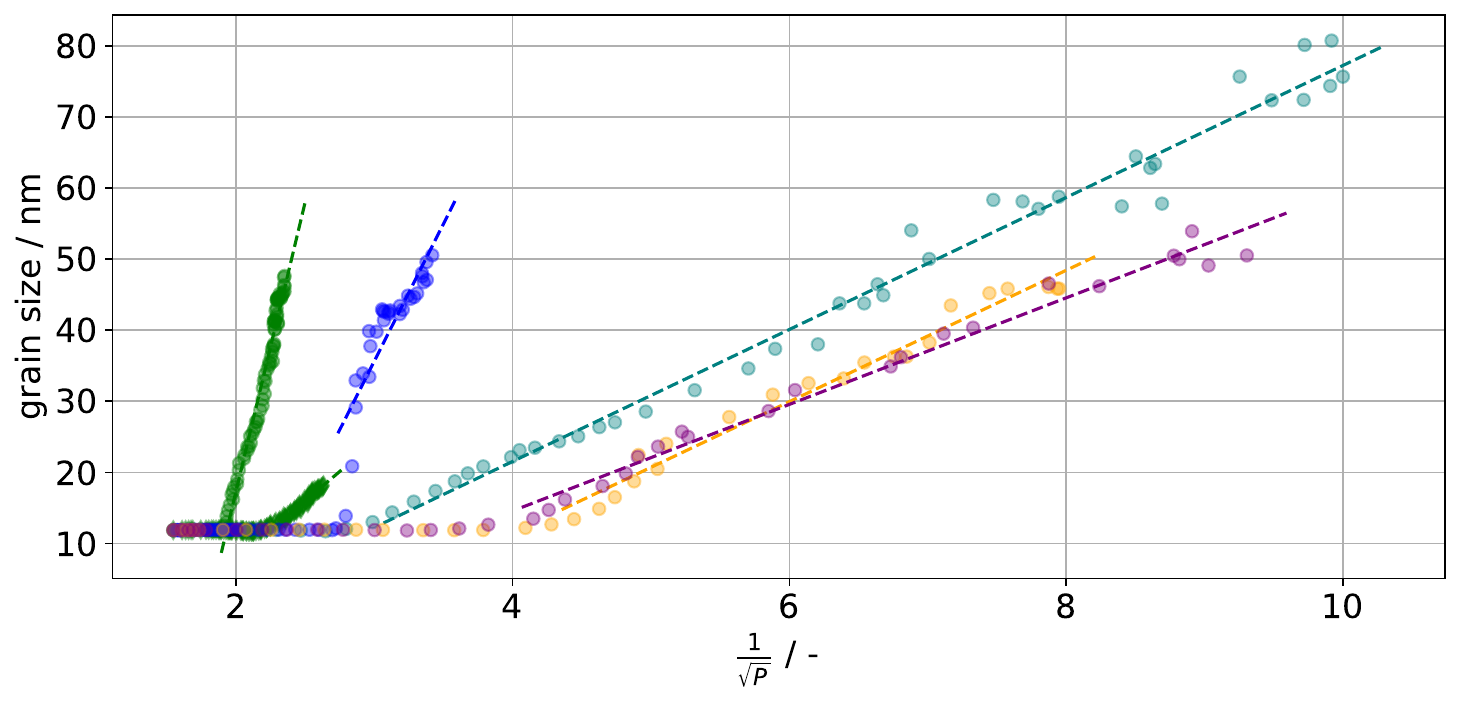}
    \caption{\cor{simulation results}}
  \end{subfigure}
    \begin{subfigure}[]{0.9\columnwidth}
    \includegraphics[width=\textwidth]{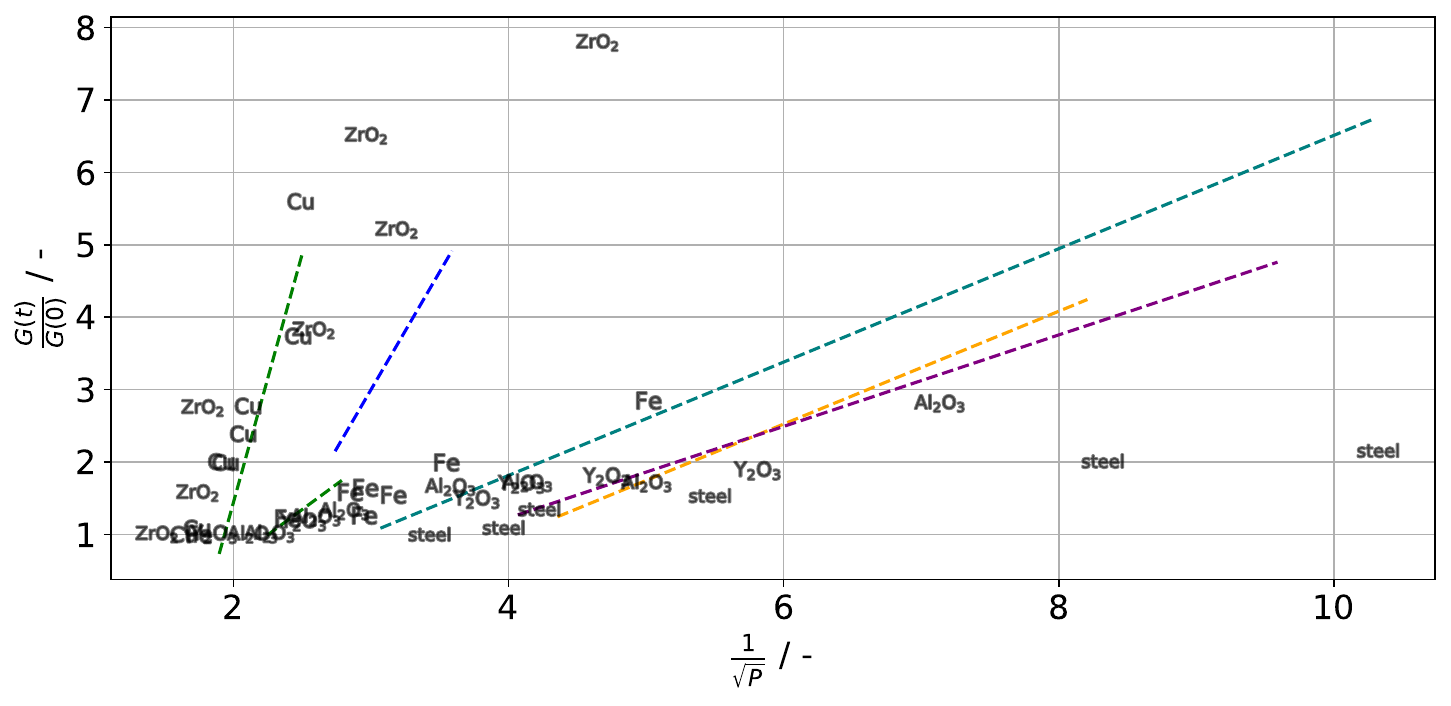}
    \caption{\cor{comparison against normalized experimental data}}
    \label{fig:exp-comparison}
  \end{subfigure}
 \caption{The inverse square root of porosity $P$ is plotted against the grain size $G$.
 Once grain growth starts, a rather linear relationship is observed; the dashed lines indicate fits to \cor{simulation} data with $G>\SI{13}{nm}$ to account for this.
 \cor{In (b), only the fits together with normalized experimental data is depicted.
 The data are taken from the overview article \cite{German2016} figs. 8 (Cu) and 9 (Fe, $\mathrm{ZrO_2}$, steel, $\mathrm{Al_2O_3}$) and \cite{Granger2007} ($\mathrm{Y_2O_3}$).
 The grain size is normalized to the grain size of the first data point.}
}
 \label{fig:gs-porosity}
\end{figure}

Based on these observations, one may conclude that below a certain ratio of $\frac{D_s}{D_{gb}}$, densification will proceed unhampered by surface diffusion.
The coarse spacing of the ratios of the present investigation does not allow a particularly accurate estimate of the critical ratio, but it should be in the interval $[1,10]$, i.e. GB diffusion should at worst be only a magnitude slower than surface diffusion.
Otherwise surface diffusion will inevitably account for major parts of neck growth and hence reduce the achievable density.
Given that surface diffusion and GB diffusion are similar in their grain size dependence following Herring's scaling law, this should also extend to larger particle sizes and when grain growth occurs.
At some point however volume diffusion becomes, in terms of total mass transported, relevant w.r.t. the interfacial fluxes as it is less affected by grain size.

Let us close the classical consideration by verifying the experimentally observed linear relationship between grain size and the inverse square root of porosity $G \propto \frac{1}{\sqrt{P}}$ \cite{German2016} with $P = 1-\rho$.
This relationship is plotted for the simulations which exhibited grain growth ($G>\SI{13}{nm}$ at any point) in \cref{fig:gs-porosity} and shows a quite linear character once grain growth has started.
This is quantified with linear fits, whose coefficient of determination is generally $R^2 > 0.97$, except for \DgbrDsrc{$D_{gb} = 0.1D_{gb,0}$ $D_{s} = 0.01D_{s,0}$} ($R^2\approx0.75$) with its stagnating grain growth.
The scatter at high densities $\rho = 98\% \sim \frac{1}{\sqrt{P}}\approx 7$ is due to the density measurement being affected by surface roughness of the green body, which is effectively counted as open porosity.
However, at this stage there are no pore channels left on the green body's surface, hence any ``porosity'' counted near the surface is a measurement error.
\cor{Furthermore, by normalizing experimental data available in the literature with the initial grain size, it is possible to compare the present results directly with experimental results.
This is shown in \cref{fig:exp-comparison} and a quite good agreement between the experiments for copper and the simulation with both reduced grain boundary diffusion and mobile GBs is observed.
The remaining experimental data points also show that, in terms of the sintering trajectory, different material systems can be approximated by the present model.
}

\begin{figure}[t!]
  \begin{subfigure}[]{0.9\columnwidth}
   
    \includegraphics[width=\textwidth]{images/legend}
  \end{subfigure}
  \begin{subfigure}[]{0.9\columnwidth}
 \includegraphics[width=\columnwidth]{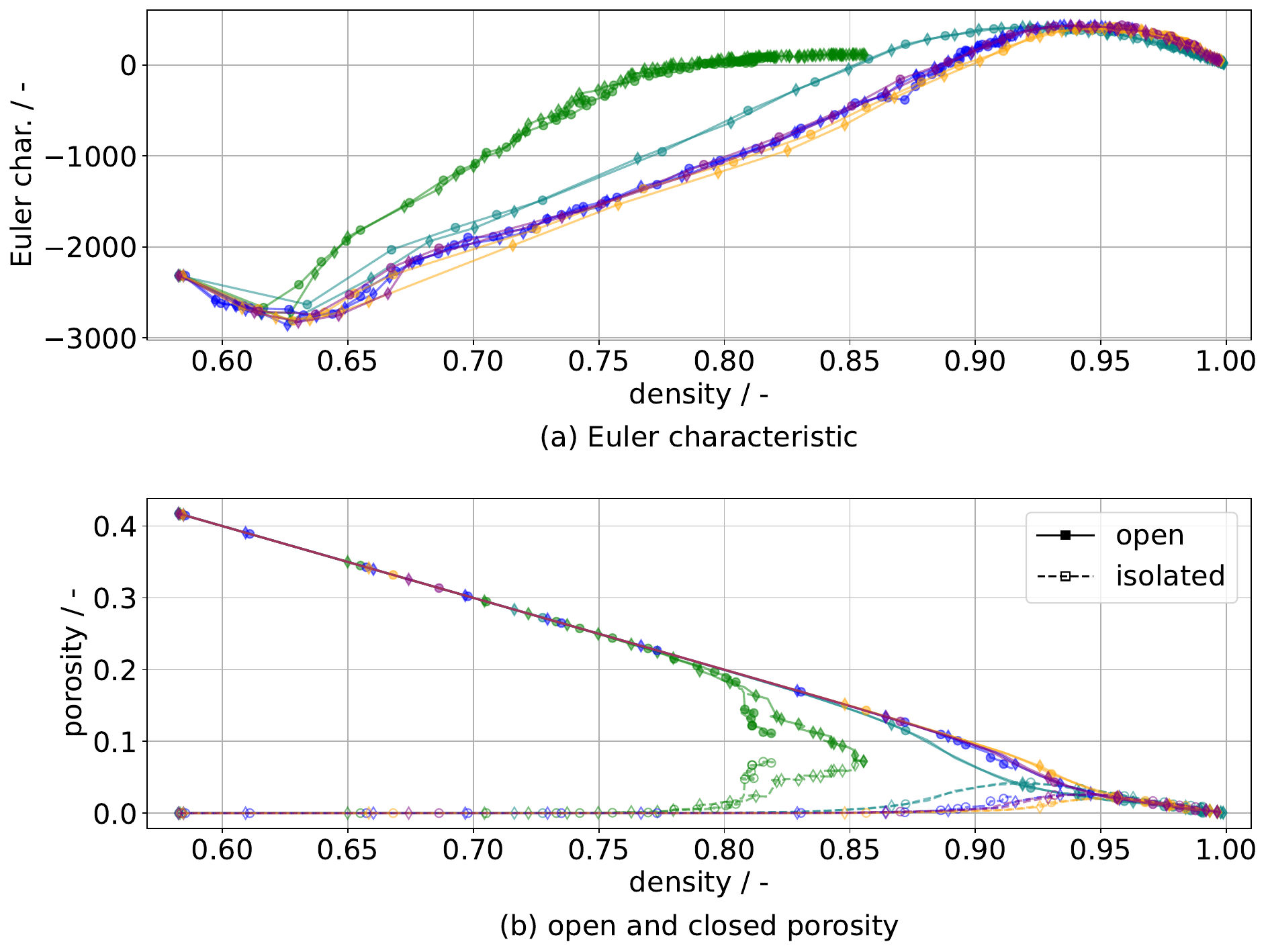}
 \end{subfigure}
 \caption{The Euler characteristic $\chi$ shows qualitatively similar behaviour compared to experimental data \cite{Wakai2014,Okuma2017}.
 Its evolution is influenced by both GB and surface diffusion, with little effect from grain growth.
 \cor{The maximum of $\chi$ is observed to occur at similar densities as the crossover of open and closed porosity shown in (b).}
 }
 \label{fig:euler-density-porosity}
\end{figure}

Besides the classical quantities of density and grain size, the microstructure can also be characterized via its shape.
An integral approach to this characterization is the Euler characteristic $\chi$ of the pore space, which can be used to distinguish stages of sintering \cite{Wakai2014,Okuma2017}:
The initial stage is characterized by a negative and decreasing value, with the intermediate stage continuously increasing the value up to a positive maximum value.
In the final stage, the characteristic decreases towards zero for infinite time, representing a fully dense structure.
The Euler characteristic is plotted over density in \cref{fig:euler-density-porosity} for the present simulation set and good accordance to the expectations is observed.
Note that grain growth has little effect on the evolution of the characteristic, since the symbols generally overlap.
It is however influenced by GB diffusion via densification speed and by surface diffusion because it determines the time scale of pore evolution.
In order to verify the suggestion of \cite{Okuma2017} that the maximum of $\chi$ indicates the final sintering stage, the bottom part also shows open and closed porosity separately over density.
Assuming that the start of the final stage is given by the equality of isolated and open porosity, the suggestion is confirmed for the present simulations.
Furthermore, it can be observed that isolated porosity is formed more slowly with \Dsrc{lower} \Dsrrc{surface diffusion}.
This is simply due to the pore instability being mostly limited by surface transport.
Fast densification also has an apparent pore instability suppressing effect, since \Dsrrc{$D_s=0.01D_{s,0}$ (orange)} has generally less isolated porosity than \DgbrDsrc{$D_{gb}=0.1D_{gb,0} D_s=0.01D_{s,0}$ (blue)} at the same density.
The likely origin of this is that densification progresses rapidly enough that the time to destabilization isn't reached before the pores are eliminated.

An approach which allows a more granular shape description is the calculation of shape factors characterizing individual objects.
The factors employed are those of MacSleyne et al. \cite{MacSleyne2008} employing second moments of the mass distribution of the object to characterize its shape.
These have three invariants $\widetilde{\Omega}_i$, $i \in \{1,2,3\}$ , which are characteristic for a fixed shape e.g. a sphere, a tetrahedron or the truncated octahedron (tetrakaidecahedron) suggested by Coble \cite{Coble1961} for modelling the intermediate and final stage grain shape.
The invariants can be normalized to a specific shape $s$, $\Omega_i =\frac{ \widetilde{\Omega_i}}{\widetilde{\Omega}_{i,s}}$, and thus the distance from this shape is related to how far from the value of $1$ the normalized invariant $\Omega_i$ is.
The sphere is chosen as a reference shape and the normalized variants are collected in a vector $\Omega$.
In order to summarize the invariants, we introduce the lumped invariant employing the Euclidean norm $\Omega_E = \frac{||\Omega||}{\sqrt{3}}$ of the vector $\Omega$.
The factor of $\sqrt{3}$ stems from the use of the Euclidean norm, such that $\Omega_E = 1$ still represents a sphere; the value itself can be interpreted in the same way as the components $\Omega_i$.
The invariants are separately calculated for the grains, isolated pores and detached pores, with the detached pores forming a subset of the isolated pores. Hence if only detached pores remain, their invariants will be the same as for the isolated pores.

\begin{figure}[b!]
\centering
\begin{subfigure}[]{0.98\columnwidth}
        \includegraphics[width=\textwidth]{images/legend}
\end{subfigure}
      \begin{subfigure}[]{0.49\columnwidth}
    \centering
        \includegraphics[width=\textwidth]{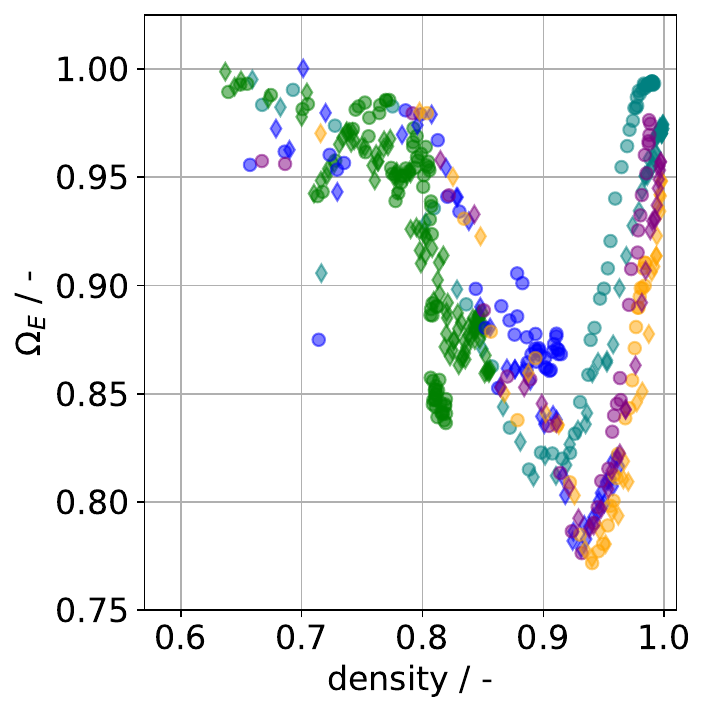}
        \caption{isolated porosity} 
        \label{fig:om3_isolated}
    \end{subfigure}
        \begin{subfigure}[]{0.49\columnwidth}
    \centering
        \includegraphics[width=\textwidth]{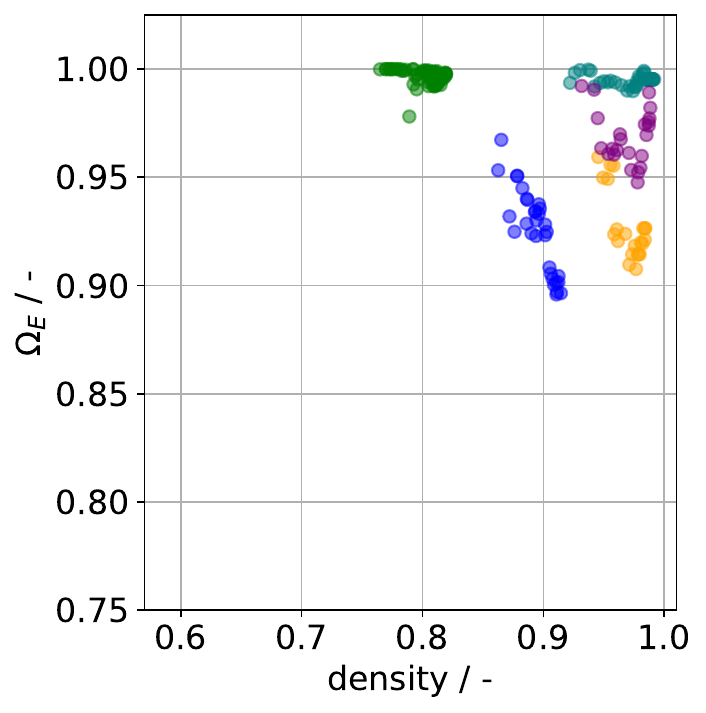}
        \caption{detached porosity} 
        \label{fig:om3_detached}
    \end{subfigure}
    \begin{subfigure}[]{0.98\columnwidth}
        \centering
        \includegraphics[width=\textwidth]{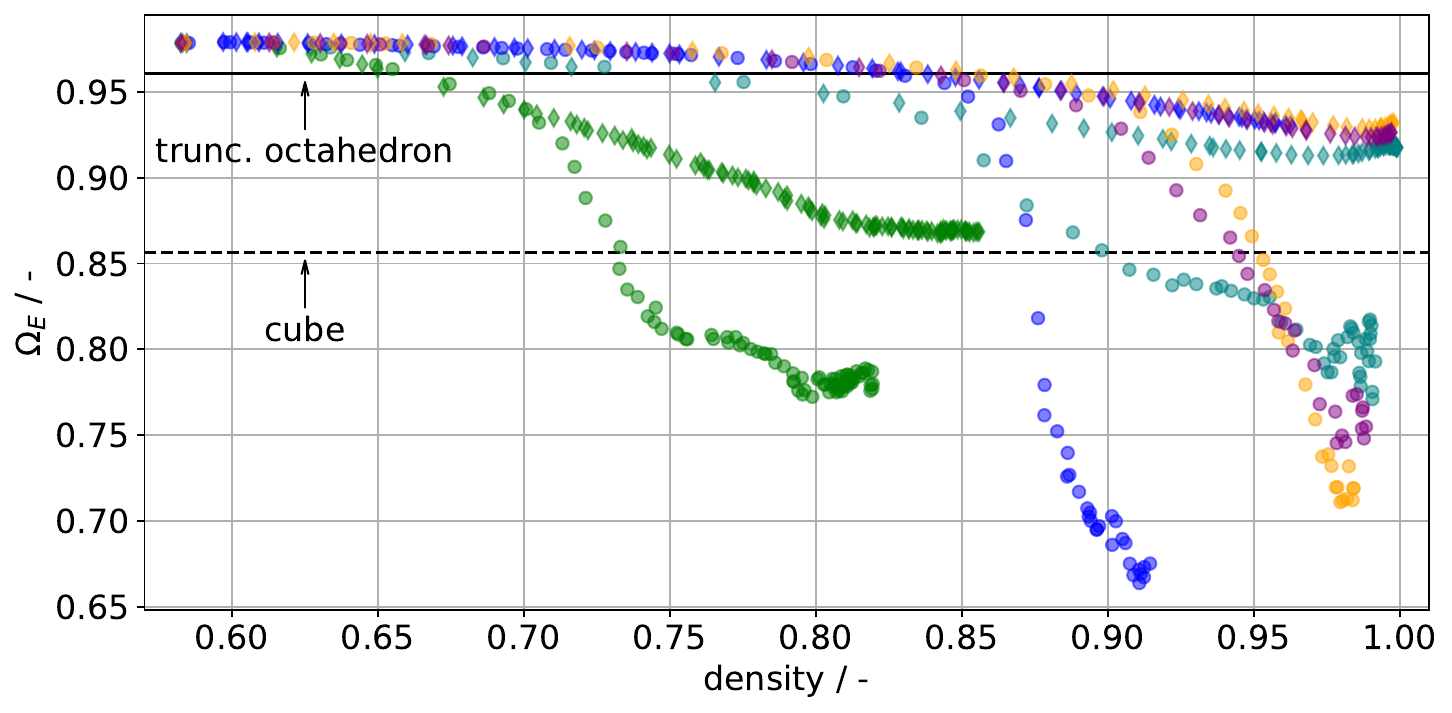}
        \caption{grains} 
        \label{fig:om3_grain}
    \end{subfigure}
    \caption{The lumped invariant $\Omega_E = \frac{||\Omega||}{\sqrt{3}}$ of the normalized invariant vector $\Omega$ is plotted over density for all simulations, with the structures of isolated and detached porosity as well as the grains being shown separately.
    Given sufficient densification, the porosity tends to approximate spheres, i.e. $\Omega_E=1$.
    The steady invariants for the grains depend on densification rate and grain mobility, with the latter influence likely due to finite size effects.
    }
    \label{fig:invariants}
\end{figure}

The mean of $\Omega_E$  is plotted over density in \cref{fig:invariants} for the separate structures.
The standard deviation was also calculated, but in general it is quite large and hence would obscure the evolution.
It was ensured that there are at least 5 objects of which the mean is calculated, as otherwise the initial trajectory is dominated by small, short-lived pores for the pores.

Focus first on the evolution of the isolated and detached pores:
At high densities, reached by simulations with high GB diffusivity, both tend to $\Omega_E = 1$, which is consistent with curvature minimization.
The transitory period shows a non-monotonic behaviour, likely attributable to isolated networks of pores first forming, then splitting into single pores via Plateau-Rayleigh-like instabilities.
These instabilities are likely induced via grain boundaries since this is typically faster than instability growth from perturbations. \cite{Hussein2022}.
Since the timescale for this instability is dependent on the surface diffusion, the convergence to a spherical shape is also slower for \Dsrc{reduced} \Dsrrc{surface diffusion}.
High grain mobility allows pores to detach and afterwards spheroidize within the grains.
This causes the invariant for the simulations with high grain mobility (circles) to generally be closer to that of the sphere than for those without, since pores on grain boundaries deform to reach the proper dihedral angle.
In the slowly densifying cases (\Dgbrc{low} \DgbrDsrc{GB diffusivity}), the isolated pore networks still remain and tend to dominate the shape factor.
Note that an entire pore network can become detached from the GB network, which significantly slows down its destabilization to isolated spheres.
Hence apparently isolated porosity in 2D micrographs might very well be the same pore.

Focus next on the evolution of the grain structures:
From the initial somewhat spherical shape, a mostly monotonic decrease is observed, with the speed being related to both interfacial diffusivities and the grain mobility.
The dashed black line indicates the cube's invariant, with the solid black line indicating the truncated octahedron's invariant.
All simulations pass the truncated octahedron's invariant, but do not converge towards it.
One might argue that pore-laden structures have a different invariant compared to the pore-free structure.
But even for the base case without grain growth, in which only few pores remain at the end, a value significantly different, i.e. $\approx \frac{1}{3}$ of the distance between a sphere and a cube, from that of the truncated octahedron is observed.
Experimental observation of the annealing of dense ($\approx99\%$) strontium titanate \cite{Trenkle2020} also indicate that real grains do not show the invariants of the truncated octahedron, though these of course also have effects from crystalline anisotropy.
Together with the evolution of grain coordination number before the final sintering stage \cite{German2014b,Seiz2023b}, this suggests that the assumption of truncated octahedra being representative of the grain structure during intermediate stage sintering should be dropped.
However, the present results do not allow a constructive suggestion for a replacement shape.
The generation of shapes from invariants, while possible, usually requires invariants of higher order moments as shown in \cite{Novotni2004}; furthermore, these shapes are likely not as analytically tractable as the truncated octahedron.
Hence while the truncated octahedron is not a satisfactory approximation for the grain shape during intermediate stage sintering, it can still be used in geometric modelling while being aware that prefactors derived from it will likely be wrong.\\
As a final note, the invariants for simulations with unsuppressed grain growth will be affected by the initial green body shape in their later stages.
This is due to the few leftover grains approximating the original green body shape.
Therefore, the grain invariants' values in the later stages with unsuppressed grain growth should not be taken at face value.
The plots of the unlumped invariants, generally showing similar trends, are deposited with the supplementary material.

\subsection*{Sintering of freeze-cast structures}
Freeze-casting \cite{Deville2017c} is a novel process for the production of porous materials as well as near net shape casting.
A suspension of a liquid, usually water and hence assumed thus, and a target material is mixed and frozen, resulting in a microstructure of the target material and ice.
The ice is sublimated next, leaving a porous structure which is then usually sintered, completing the freeze-casting process chain.
A previously developed model for the phase-field simulation of the freezing part of freeze-casting \cite{Seiz2021b} is employed to generate a three-dimensional freeze-cast structure.
The volume fraction of target material is known for each cell of the domain and is employed to generate a sphere packing approximate the freeze-cast body.
Where a sufficient volume fraction is reached, spheres of $r = \SI{8}{nm}$  are placed into a domain, approximating the continuous volume fraction field with a discrete packing of spheres.
The freeze-casting simulation itself employed particles of radius $\SI{250}{nm}$, with the change in sphere radius being for computational efficiency, as smaller particles will sinter faster; this does mean however that a smaller version of the actual structure is sintered next.
The space of low volume fraction is left empty, producing macropores, and thus a rough approximation of the sublimation process is achieved.
Finally, the resulting sphere packing is computationally sintered with the present model to completely simulate the process chain of freeze-casting for the first time ever.

\begin{figure}
\centering
\includegraphics[height=0.6\textheight]{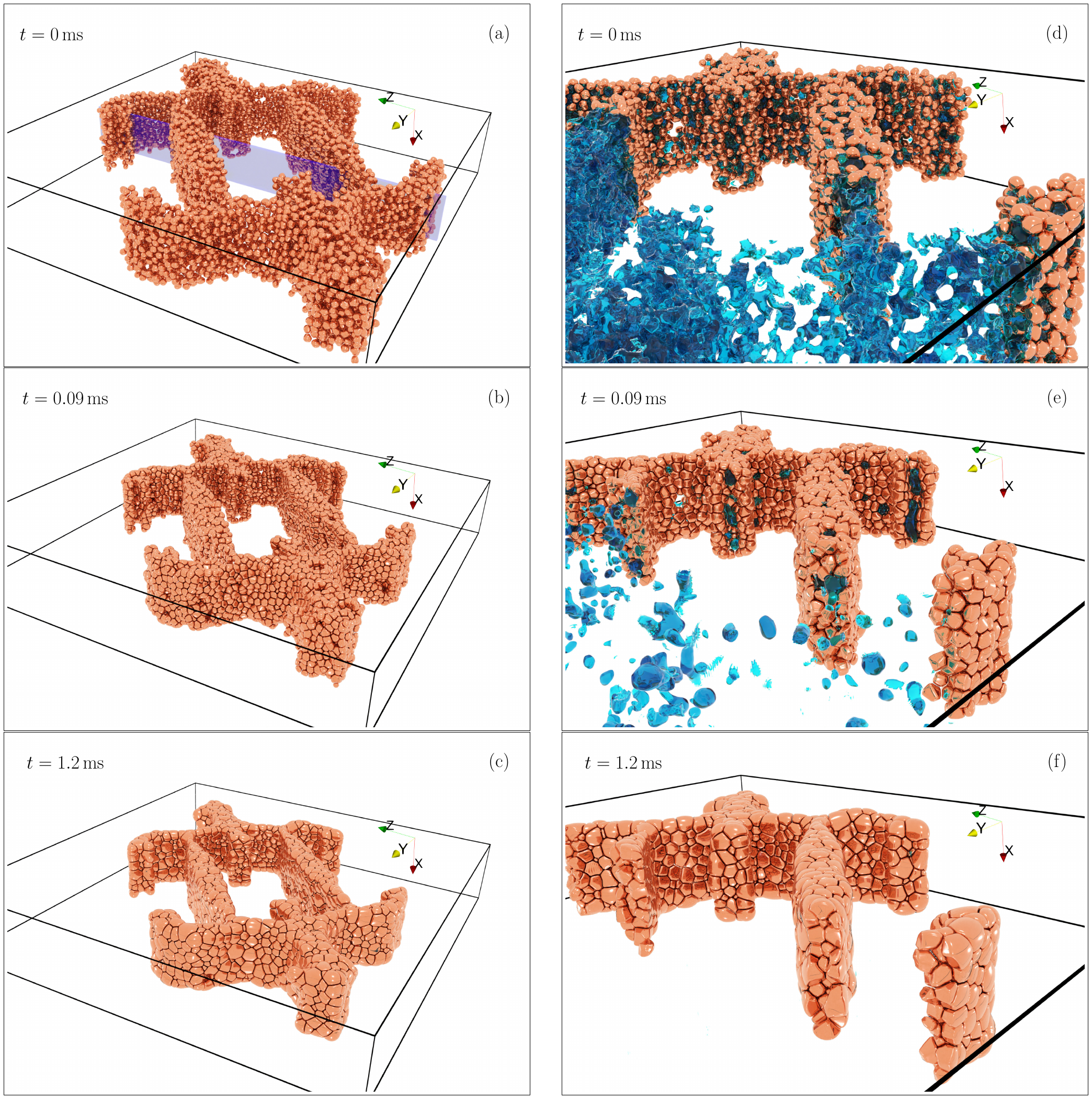}
 \caption{A freeze-cast structure with a macropore undergoing sintering.
 \cor{The freezing direction was parallel to the $X$ direction shown in the figures.}
 The structure itself shrinks while keeping the macropore intact.
 \cor{Panels (a-c) show the full body, whereas panels (d-f) show a close-up for a fracture surface normal to the $Y$ direction, also indicated by the blue plane at $t=\SI{0}{ms}$, including a rough approximation of the porosity rendered with a water-like appearance.}
 It can easily be seen that the porosity within the walls changes substantially between $t=\SI{0}{ms}$ and $t=\SI{0.09}{ms}$, with many grains already having eliminated their surrounding porosity.
}
 \label{fig:fc-pics}
\end{figure}

The resulting microstructural evolution over time is shown in \cref{fig:fc-pics}.
On the left the entire structure is shown, with grains being rendered as a copper-like material, since the materials parameters of \cref{tab:params} approximate copper, and the dark lines within the structure represent grain boundaries and higher order junctions.
As can be seen, even a complex, inhomogeneous structure can be sintered with the present model.
Furthermore, some grain growth has occurred as would be expected from the low grain size.
At the end of the simulation run, the particulate region is completely dense, with no isolated porosity remaining and no pore channels penetrating through the structure.
On the right a fracture surface is shown together with a rough approximation of the porosity, rendered with a water-like apperance.
This porosity is continuously reduced, with many grains already having eliminated their surrounding porosity at $t=\SI{0.09}{ms}$, which limits their further contribution to densification.
Note that almost no vertical motion is evident between $t=\SI{0.09}{ms}$ and $t=\SI{1.2}{ms}$.
Densification in this direction has stopped at about $t=\SI{0.09}{ms}$  and only the directions normal to it continue to densify.
The surface in this direction also tends to have fewer grains, allowing these to bulge out more and thereby influence the strain measurement.

The shrinkage is characterized with the strain in each spatial dimension, which is found to be anisotropic:
In the freezing direction, upwards and parallel to the macropore, a shrinkage of about 10 to 13\% is observed, but in the directions normal to it, a shrinkage of 16 to 19\% is observed.
The strain range is due to inhomogeneous strain measurements.
Experimental evidence of anisotropic shrinkage after freeze-casting exists \cite{Farhangdoust2013,Lichtner2018}, but the experiments disagreed on which direction shrinks less.
Farhangdoust et al.\cite{Farhangdoust2013} observed that the freezing direction shrunk less, comparable to the present results.
However, as Lichtner et al.\cite{Lichtner2018} note, Farhangdoust et al. may not have removed the initial, isotropic structure generated by freeze-casting and neither the continuous skin formed on the outside of a freeze-cast cylinder.
Lichtner et al. investigated the anisotropic shrinkage with experiments and the discrete element method, experimentally and simulatively observing less shrinkage in the plane normal to the freezing direction.
Following Olevsky\cite{Olevsky1998} it was argued that the anisotropic shape of the pores induces anisotropic shrinkage.
The reason for this is that the sintering potential is proportional to curvature, and if the curvature is anisotropic, so is the sintering potential.
Hence directions which exhibit a larger curvature will densify faster than others with smaller curvature.
It should be said however that the macropores' macroscopic curvature is low and it is not clear whether it can cause the observed magnitude of shrinkage anisotropy.
Another reason stated in the discussion of \cite{Lichtner2018} was that particles within the walls will sinter isotropically, as their contacts are isotropically distributed.
Particles on the walls' surface have missing contacts and hence will tend to move anisotropically.
This disregards however that the sintering potential can also be distributed inhomogeneously and anisotropically.

\begin{figure}
\centering
 \includegraphics[width=\columnwidth]{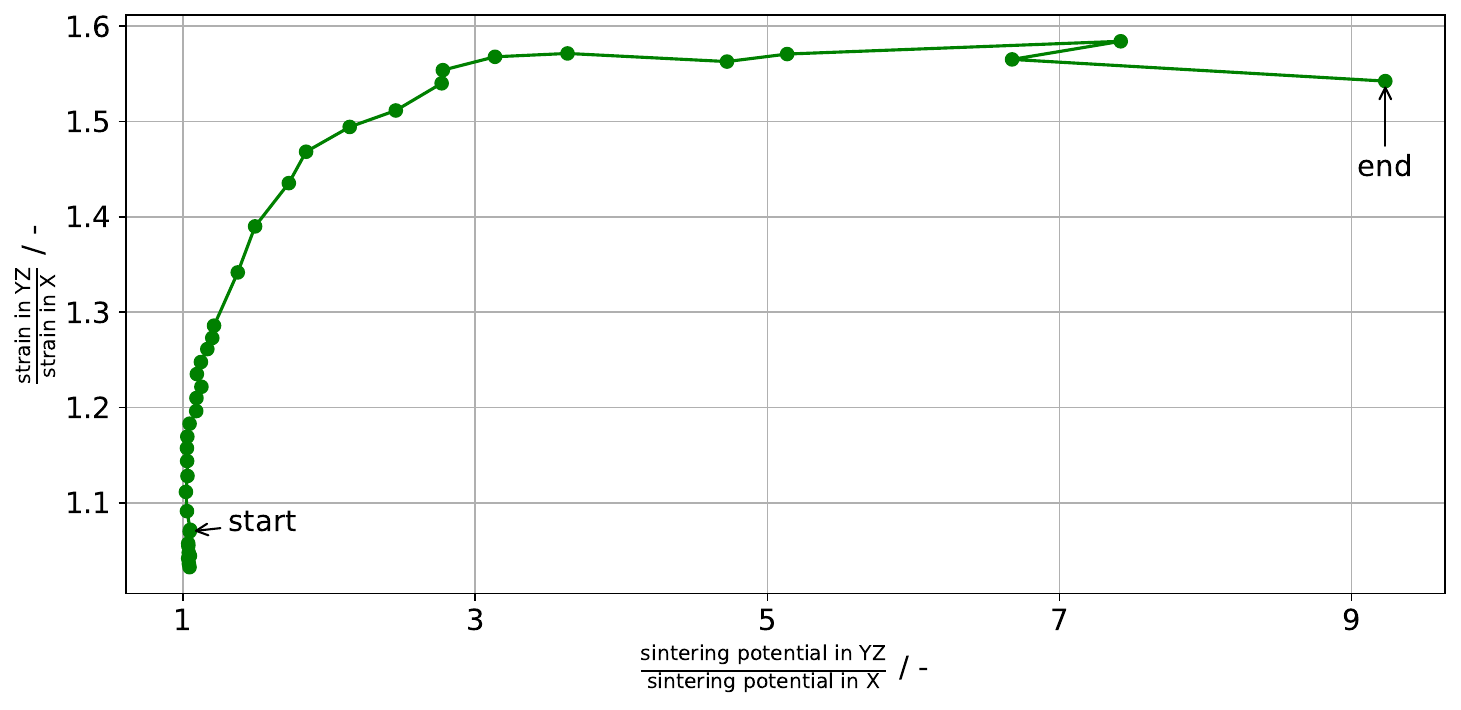}
 \caption{Anisotropy factor $f(\Pi) = \frac{\Pi_{YZ}}{\Pi_{X}}$ for the properties $\Pi$ of strain and sintering potential.}
 \label{fig:aniso}
\end{figure}

The present results can be explained by analyzing the anisotropy of the sintering potential, represented by how much GBs deviate from their equilibrium concentration within the simulation.
Define the anisotropy factor of a property $\Pi$ as $f(\Pi) = \frac{\Pi_{YZ}}{\Pi_{X}}$, i.e. simply the ratio of the property in the $YZ$ (average over both $Y$ and $Z$ directions) plane, normal to the freezing direction, to the property in the freezing direction $X$.
The anisotropy factor of both the strain and sintering potential are plotted against each other in \cref{fig:aniso}.
Two regimes are evident here:
Initially, there is a slight anisotropy in the sintering potential, which is sufficient to cause the strain to become increasingly anisotropic.
Although the anisotropy is small ($\approx 1.05$), this small anisotropy causes about half of the observed anisotropy in the strain.
This is due to the high driving forces for densification at this early stage.
The anisotropy in the sintering potential increases eventually, due to the sintering potential being connected with the chemical potential (stress) on a particle's surface within the phase-field model.
As the structure densifies, more and more grains completely lose their connection to the surface and become embedded in a surrounding grain matrix; this can easily be seen on the right of \cref{fig:fc-pics}.
Once this happens, the model assumes that there is no additional stress due to curvature and correspondingly reduces the sintering potential of the associated grain boundaries.
While this occurs isotropically, the average sintering potential in a direction is dependent on both such embedded grains and surface grains.
A simple model of this effect could read as
\begin{align*}
 \sigma^{d} = \frac{N_v \sigma_{iso} + N^{d}_s \sigma^{d}_s}{N_v + N^{d}_s}
\end{align*}
i.e. the average sintering potential $\sigma^{d}$ in a direction $d$ is the addition of both a volume contribution $N_v\sigma_{iso}$, acting isotropically, and a surface contribution $N^{d}_s \sigma^{d}_s$, potentially inducing anisotropy.
If $N_v \gg N^{d}_s$, isotropic behavior is observed, which was the case in the previous sections; though if the anisotropic contribution is the same in each direction, this would also result in net isotropic behavior.
In the present geometry however, $N_v \approx 2N^{d}_s$ in the $YZ$ plane since the walls are on average about 6 particles thin.
In contrast to this, $N_v \approx 5N^{d}_s$ in the $X$ direction (12 particles) and hence this direction will be affected less by the surface grains, especially compared to the $Y$ and $Z$ directions.
Thus, the observed anisotropy of the sintering potential might be due to the phase-field model giving an unwarranted extra weight to particles with exposed surfaces by virtue of their increased sintering potential $\sigma^{d}_s$.
Finally, the increase in the anisotropy of the sintering potential at roughly constant strain anisotropy is due to little to no porosity remaining.
This correspondingly magnifies the aforementioned effects, though given that little to no densification is possible anymore, no further increase in strain anisotropy is observed.
The non-monotonic behaviour in this final region is mostly due to measurement errors via grain growth, since densification has effectively stopped.

Three other confounding factors exist:
First, the present geometry only contains a single macropore open to the surrounding vacuum, raising questions of representativeness.
Second, within the present phase-field model it was observed that linear chains of particles only achieve particle count independent strain evolution starting from about 16 particles in the chain\cite{Seiz2023b}.
Up to that point, increasing the number of particles in the chain decreased the strain rate monotonically.
The strain rate ratio between a four and sixteen particle chain, roughly comparable to how thick the freeze-cast structures is in the $YZ$ and $X$ directions, is on average about $1.17$.
This is in rough agreement with the initial increase of strain anisotropy up to about a factor of $\approx1.2$, after which the increase in sintering potential anisotropy becomes the dominant influence on the strain anisotropy.
The change in strain rate between chains of different particle counts was explained via the differing sintering potentials of the chain's end particles.
Since their shape evolution is only restricted by one grain boundary, they will generally have an average surface chemical potential which differs from the inner particles.
Since there are many thin sections in the freeze-cast structure, this effect is present in freeze-casting sintering simulation as well, and is the likely origin of the sintering potential anisotropy.
The final confounding factor is that some grain growth ($\approx$ 10\% average grain size increase) occurs.
While for the present results it does not result in desintering \cite{Sudre1992}, the green body resulting from freeze-casting is generally liable to experience this phenomenon --- the bridges spanning the macropores are similar to the bridges spanning cracks in the classic work of Sudre and Lange \cite{Sudre1992}.
Since desintering would generally reduce the connectivity normal to the freezing direction, it would support Lichtner et al's \cite{Lichtner2018} experimental results.
In total, while the present results are quite encouraging for relatively thick structures, further research and modelling work should be done on the calculation of the sintering potential within the model.

\section*{Conclusion}
\cor{It could be shown that pressure-assisted sintering and Coble creep can be effectively modelled by the presented phase-field model of sintering.
This was shown by comparing the dependence of densification and strain rate on the applied stress magnitude, with the same scaling being found as predicted by theory\cite{Coble1963,Coble1970} and found in experiments\cite{Harmer1980}.
Furthermore, the coupled phenomena of grain growth and densification were investigated by systematic parameter variations.
A high grain boundary diffusion and a low surface diffusion are generally found to increase the densification rate.
As expected, once grain growth occurs formerly isolated porosity can detach from grain boundaries and stays stable within the grains.
This causes an eventual stagnation of densification prior to full density.
The extent of the difference between this stagnated density and full density depends on the densification speed relative to the speed of grain growth:
If grain growth occurs similarly fast as densification, 10 to 20 percent of porosity is observed to remain in the green body, whereas if densification occurs much faster than grain growth, only only about one percent of porosity remains.
These simulations also allow comparing sintering trajectories with experimental data, showing good agreement for some experimental systems even though the parameters were chosen without prior reference to the experimental data.
Furthermore, the grain morphology during the sintering process was analyzed and found to generally not conform to the classically assumed shape of a truncated octahedron.
Finally, the sintering process was applied to a more complicated geometry than a cube, namely one which was generated by the freeze-casting process.
Anisotropic shrinkage was found, with a comparison against literature data on anisotropic shrinkage of freeze-cast structures suggesting further possible model improvements.}

In total, the results show the versatility and applicability of the phase-field method to the sintering process.
Given accurate information about diffusivities and mobilities, it can be employed for quantitative predictions, since all the necessary qualitative features of the sintering process are present.
Especially for novel high heating rate sintering processes, the control of the temperature-time curve is critical to avoid grain growth following densification, with experiments being costly and time intensive.
Simulations could partially replace these and allow the determination of desirable temperature profiles.
However, finding a consistent data basis in terms of diffusivities and mobilities for a single material is still challenging, as these are not only dependent on temperature but also on the grain boundary and surface characters, with impurities as well as dopants also playing significant roles.

Future modelling work will investigate variations in how the sintering potential is calculated.
Separate from this, a non-conservative evolution of vacancies coupled with other species will be considered.
This will enable including gas pressure, and hence pore stabilization, via another conserved species.
The rigid-body assumption for individual grains can be relaxed by solving a momentum transport equation while accounting for local strain rates induced by e.g. vacancy absorption.
This would allow both volume sinks of vacancies, e.g. dislocations, as well as using an orientation field for resolving grains of different orientation, which in turn could allow the inclusion of atomistically informed interface properties.

\section*{Methods}

\paragraph{Phase-field model}
\label{sec:model}
The model employed in this paper is based on the recent works \cite{Hoetzer2019,Seiz2023a,Seiz2023b}.
\cor{In the supplementary material a derivation of the model is given together with the necessary assumptions.
The final results are summarized here.
}
A coupled system of partial differential equations
\begin{align}
\pdiff{\phi_\alpha}{t} + \nabla \cdot (\vec{v}_\alpha(x)\phi_\alpha) &= \frac{1}{\tau(\phi) \epsilon } \Big[
-\epsilon  \left(\pdiff{a(\phi,\nabla\phi)}{\phia} - \nabla \cdot \frac{\partial a(\phi,\nabla\phi)}{\partial\nabla\phia} \right) \label{eq:phi}  \\
&-\frac{1}{\epsilon}  \pdiff{w(\phi)}{\phia} \nonumber
- \sum^{N-1}_{\beta=0} \psi_{\beta}(\mu, T) \pdiff{h_{\beta}(\phi)}{\phia} \Big] - \lambda \nonumber,\\
\pdiff{\mu}{t} &=\left[ \sum_{\alpha=0}^{N-1} \ha(\phi) \left(\frac{\partial c_\alpha(\mu, T)}{\partial \mu} \right)\right]^{-1}  \nonumber \\
&\Biggl( \nabla \cdot \Big(M(\phi,\mu,T)\grad{\mu} - \vec{v}(x)c \Big) 
- \sum_{\alpha=0}^{N-1} c_\alpha(\mu,T) \frac{\partial h_\alpha(\phi)}{\partial t} \Biggr). \label{eq:mu} 
\end{align}
is solved, with the $N$ values $\phi_\alpha$ representing the surrounding vapor $\phi_0 = \phi_\Vap{}$ as well as copper grains of arbitrary orientation $\phi_\alpha, \alpha > 0$.
The evolution of the copper concentration $c$ is indirectly calculated with the chemical potential $\mu$, employing the grand potential formalism\cite{Plapp2011} to decouple bulk free energies from interfacial terms.
The concentration $1-c$ can be interpreted to be the vacancy concentration, which is low in the solid and high in the surrounding vapor.
The meanings of the remaining terms are described in detail in \cite{Hoetzer2019,Seiz2023a}.
On all fields no-flux conditions are employed as boundary conditions.

The equations include advective terms which are calculated based on a model inspired by molecular dynamics simulations:\cite{Seiz2023b}
\begin{align}
 v_\alpha &= \frac{u_\alpha}{\Delta t} \\
 \Delta \vec{u}_{\alpha\beta} &= \frac{1}{V_{\alpha\beta}} \int_{GB} 4 \phi_\alpha \phi_\beta \frac{\Omega}{A_{\alpha\beta}}\frac{N_a}{V_m} \Delta c^{gb}_{\alpha\beta}  \vec{n}_{\alpha\beta} dV \label{eq:dispjump}  \\
 \pdiff{c^{gb}}{t} &= -\frac{c-c^{gb}_{eq}(S)}{t_r} \label{eq:vacabs} \\
 \contact{} u &= \Delta u \label{eq:contact}
\end{align}
in which the instantaneous velocity follows from the instantaneous displacement.
The displacement jump $\Delta u$ across a GB of area $A_{\alpha\beta}$ is based on how many vacancies $\propto \int_{GB} \Delta c^{gb}_{\alpha\beta} dV$ were absorbed on a single $\alpha\beta$ GB; this is scaled by the atomic volume $\Omega$.
The expression $c-c^{gb}_{eq}(S)$ can also be thought of as the sintering potential; once it vanishes, densification via advection stops.
The factor $\frac{N_a}{V_m}$ with Avogadro's constant $N_a$ and the molar volume $V_m$ accounts for the conversion between units of concentration $c$ (mole fraction) and number density $n$.
The absorption rate is based on a relaxation to an equilibrium vacancy concentration $c^{gb}_{eq}(S)$ on a GB.
This depends on the simulation state $S$ in order to account for the Gibbs-Thomson effect.
It is approximated via the concentration of a bulk $\alpha$ grain $c_\alpha$ at its average surface chemical potential $\mu_s$\cite{Seiz2023a}.
Finally, the displacement jumps are connected to individual grain displacements $u$ in a matrix equation\cite{Seiz2023b}.
\cor{For efficient parallelization, a linear ansatz function for the displacements $u$ is employed.
This restricts the simulation geometry to those in which the vacancy absorption rate \cref{eq:vacabs} does not have an explicit dependence on position, but can still vary from GB to GB.
However, the model itself is capable of handling arbitrary variations in grain size or vacancy absorption rates as shown in the supplementary material.}

By applying a bias in $\mu_s$ the effect of pressure on the equilibrium vacancy concentration on grain boundaries $c^{gb}_{eq}(S)$ can be modelled.
Furthermore, the effect of uniaxial stress can be approximated by a projection onto the local grain boundary plane $\vec{n}_{\alpha\beta}$:
\begin{align}
 c^{gb}_{eq}(S) &= c_\alpha(\mu_s + \sigma(\vec{n}_{\alpha\beta}))\\
 \sigma(\vec{n}_{\alpha\beta}) &= \sigma (\vec{v}_\sigma \cdot \vec{n}_{\alpha\beta})
\end{align}
with the unit vector $\vec{v}_\sigma$ describing the direction of stress application.
This effectively presumes that grain boundaries with normals aligned to the stress direction experience the full stress, whereas those perpendicular to it experience no stress.
This modifies the vacancy absorption rate per grain boundary and depending on the sign of $\sigma$ can also cause vacancy generation.
In the case of an isotropic pressure, this projection can be skipped and the pressure $p=\sigma$ can be added directly.

\cor{This approach of adding stress is inspired by Coble's\cite{Coble1970} classical model in which the capillary pressure and external pressure are simply added together.
A more general approach would include the stress and strain fields in the energy functional (cf. the derivation within the supplementary material), coupling them with the chemical potential and thus also change the local equilibria.
Furthermore, if the stress fields are locally resolved, the effects of stress gradients on densification would also be included.
However, this requires solving for the stress evolution as well, which adds considerable additional computational effort, especially if a realistic material contrast between solid and vapor is considered.
The effect of stress is not limited to densification, but also influences grain growth:
Pressure modifies the diffusivity \cite{Mehrer2007} and therefore the grain mobility, with higher pressures typically decreasing the diffusivity.
However, at the same time, stress concentrations can lead to higher driving forces for grain growth.
}

\cor{Let us contrast this to the work of Dzepina et al.\cite{Dzepina2019}:
The model of Wang\cite{Wang2006} is extended by adding an approximate elastic energy term into the functional.
Its energy contribution is based on normal forces homogeneously acting on GBs, with the origin of the force being a deviation from the equilibrium vacancy density on the GB.
This construction allows the addition of an external force (pressure) by shifting the force generated from the vacancy disequilibrium by the external force.
This is similar to the present approach, except that by working directly in pressure (chemical potential) space there is no need to assume a spherial contact within the present model.
However, since the motion of a particle is the result of a direct-neighbours resultant force calculation, not a global one, the densification obtained by \cite{Dzepina2019} is unlikely to be homogeneous.
The origin of this effect is described in the supplementary material's model derivation.
Note that the stress-field is \emph{not} explicitly solved for in \cite{Dzepina2019}.}

The parameters of the simulations are given in \cref{tab:params}, with more details on the choices being given in \cite{Seiz2023a,Seiz2023b}.
The diffusion and interfacial energy values approximate copper at \SI{700}{K}, with the choice of inverse mobility $\tau$ being such that diffusion-controlled growth is achieved for all simulations.

\begin{table}[h]
\centering
\caption{Employed physical and numerical parameters for the simulations.}
\label{tab:params}
 \begin{tabular}{lll}

parameter   & nondim. value & physical value \\
\multicolumn{3}{c}{\textit{numerical parameters}}\\
grid spacing $\Delta x$   &    0.1  &   $\SI{1e-9}{m}$  \\
max. time step $\Delta t_{max}$   &    $\num{1.5e-05}$  & $\SI{1.5e-9}{s}$  \\
interface parameter $\epsilon$   &    $4\Delta x$  &   $\SI{4e-9}{m}$  \\
interface width $W\approx2.5\epsilon$   &    $10\Delta x$  &   $\SI{10e-9}{m}$  \\
grain boundary cutoff $\phi^{min}_{\alpha\beta}$ & 0.14 & - \\
\multicolumn{3}{c}{\textit{physical parameters}}\\
surface energy $\gamma_{v\alpha}$     & 2 & $\SI{2}{J.m^{-2}}$ \\
grain boundary energy $\gamma_{\alpha\beta}$     & 1 & $\SI{1}{J.m^{-2}}$ \\
volume diffusion $D$    & \num{1e-3}               & $\SI{1e-15}{m^2.s^{-1}}$  \\
grain boundary diffusion  $D_{gb,0}$    & 55               & $\SI{5.5e-11}{m^2.s^{-1}}$  \\
surface diffusion  $D_{s,0}$ & 169               & $\SI{1.69e-10}{m^2.s^{-1}}$   \\
physical interface width $\delta_i$ & 0.02 & $\SI{2e-10}{m}$ \\
surface inverse mobility $\tau_{gv}$ & 0.08 & $\SI{8e10}{J.s.m^{-4}}$ \\
grain boundary inverse mobility $\tau_{gb}$       & variable & variable \\
atomic volume $\Omega$ & $\num{1.22e-5}$ & $\SI{1.22e-29}{m^3}$\\
GB relaxation time $t_r$ & $\num{1e-8}$ & $\SI{1e-12}{s}$
\end{tabular}
\end{table}

\paragraph{Data evaluation methods}
The density of a green body is calculated as described in \cite{Seiz2023a} with a Monte Carlo approach of sampling inside the convex hull of the green body.
The strain is evaluated by dividing the domain into one-dimensional slabs per evaluated dimension $d$.
The slabs are positioned based on the current minimum and maximum position of the particles in that dimension, with a fixed number of slabs employed per simulation.
The grain-wise displacement $u_{d,\alpha}(0)-u_{d,\alpha}(t)$ of each grain $\alpha$ is calculated and averaged over for each grain within the slab.
With the average displacement per slab and the slab position, finite differences yield the normal strain $\epsilon_{dd}$ per slab position, which can be averaged over if homogeneous strain per dimension is assumed.
This definition will yield positive strains for a shrinking compact, consistent with the usual definition of strain in sintering.

The grain volume $V_\alpha$ and the equivalent grain size $G_\alpha = (\frac{3V_\alpha}{4\pi})^{1/3}$ easily follow from volume integrals over the phase-field domain.
Integral interfacial areas (surface $s$, grain boundary $gb$) are defined by pairwise clamped sums over the appropriate phase pairs, divided by the effective phase-field width $l=\frac{\pi^{2} \epsilon}{8}$:
\begin{align}
  \phi_{s} &= \min( [\sum_{\alpha\neq \Vap{}} 4\phi_\alpha \phi_\Vap{}], 1)\\
  \phi_{gb} &= \min( [\sum_{\alpha\neq \Vap{}} \sum_{\beta>\alpha} 4\phi_\alpha \phi_\beta], 1) \\
  A_s &= \frac{1}{l}\int \phi_s dV\\
  A_{gb} &= \frac{1}{l}\int \phi_{gb}dV
\end{align}
with the clamp being necessary as the sum of paired phase-field products is not constrained to be $\geq$1.
The field described by $\phi_{gb}$ is also used to visualize the grain boundary network.
The sintering potential in the axis directions is calculated by a $\phi_{gb}$-weighted average of $(c-c^{gb}_{eq}(S))$ with the components in each direction being based on the local grain boundary normal vector.

The package cc3d's\cite{cc3d} connected component labelling function is used to identify isolated porosity by applying it to the vapour phase-field.
This is followed by a sweep of the field:
If any cell of an isolated porosity object contains a non-zero value of $\phi_{gb}$, then it is assumed to still be attached, but detached otherwise.
After these labelling steps, the diffuse vapour field is split into an isolated and detached pore field which are used for analysis and visualization.

The central moments of second order $\mu_{pqr}$ required for the calculation of the moment invariants $\widetilde{\Omega}$\cite{MacSleyne2008} are calculated by integrating
\begin{align}
 \mu^{A}_{pqr} &= \int f(A) (x-x_A)^p (y-y_A)^q (z-z_A)^r dV
\end{align}
over the domain with $p+q+r=2$, for each object $A$ with its center of mass $(x_A, y_A, z_A)$ and  $f(A)=1$ inside of the object and zero outside.
The invariants follow directly as described in \cite{MacSleyne2008}.
The truncated octahedron's second moments are calculated with LattE integrale\cite{latte}, yielding numerical values of the invariants $\widetilde{\Omega} = (12.732, 162.10, 2063.8)$.

The Euler characteristic $\chi$ is evaluated on a triangle mesh of the vapour phase-field using its definition
\begin{align}
 \chi &=  V - E + F
\end{align}
with the number of vertices $V$, edges $E$ and faces $F$ of the triangle mesh. 

Paraview\cite{paraview} is used for three-dimensional visualization and plots are drawn by employing matplotlib\cite{matplotlib}.

\section*{Author contributions}
\textbf{Marco Seiz}: Conceptualization, Software, Methodology, Investigation, Data Curation, Validation, Visualization, Resources, Writing - original draft, Writing - review \& editing.
\textbf{Henrik Hierl}: Software, Data Curation, Validation, Visualization, Resources, Writing - review \& editing.
\textbf{Britta Nestler}: Funding acquisition, Writing - review \& editing.
\textbf{Wolfgang Rheinheimer}: Conceptualization, Writing - review \& editing.

\section*{Conflicts of interest or competing interests}
The authors declare that there are no conflicts of interest.

\section*{Data and code availability}
The code required to reproduce the present work cannot be shared publicly.
Parts of the processed data are available within the the supplementary material.
The raw data supporting the findings is available upon reasonable request from the corresponding author (Marco Seiz, marco.seiz@partner.kit.edu).

\section*{Supplementary Material}
% \todoMarco{update sup mat, new url}
%/data2/sintern/diffusionmodel/paper3_pub
% https://gitlab.kit.edu/marco.seiz/process-mat-influence-sintering
The Supplementary Material of this paper is available at \url{https://doi.org/10.5281/zenodo.8263532}.
It contains notebooks to produce most of the figures within the paper as well as some additional analysis.
Furthermore, videos of the time evolution of selected structures are contained as well.
\cor{Finally, the employed model is derived with special attention being paid to the velocity modelling.}

\section*{Acknowledgements}
This work was partially performed on the national supercomputer Hawk at the High Performance Computing Center Stuttgart (HLRS) under the grant number pace3d.
This work was partially performed on the BwUniCluster and authors acknowledge support by the state of Baden-Württemberg through bwHPC.
The authors gratefully acknowledge financial support by the Deutsche Forschungsgemeinschaft (DFG) under the grant numbers NE 822/31-1 (Gottfried-Wilhelm Leibniz prize) and RH 146/1-1 (Emmy Noether programme), and the financial support for the parallelization and code optimization by KNMFi within the programme MSE (P3T1) no. 43.31.01.

% \section*{References}
\bibliography{literatur}

\end{document}